\documentclass[10pt,journal,compsoc]{IEEEtran}

\usepackage{amsthm}
\usepackage{algorithm}
\usepackage{algpseudocode}
\usepackage{amssymb}
\usepackage{fancyvrb}
\usepackage{graphicx}
\usepackage{xcolor}
\usepackage{url}
\usepackage{subcaption}
\usepackage{mathtools}  
\usepackage[nocompress]{cite}
\usepackage{enumitem} 
\usepackage{soul}
\usepackage{tabularray} 
\usepackage{longtable}

\newcommand{\tb}[1]{\textbf{#1}}

\begin{document}

\title{Formal Definitions and Performance Comparison of Consistency Models for Parallel File Systems}

\author{Chen~Wang,~\IEEEmembership{Member,~IEEE,}
        Kathryn~Mohror,~\IEEEmembership{Member,~IEEE,}
        and~Marc~Snir,~\IEEEmembership{Fellow,~IEEE}%
\IEEEcompsocitemizethanks{
    \IEEEcompsocthanksitem Chen Wang and Kathryn Mohror are with Lawrence Livermore National Laboratory. E-mail: \{wang116, mohror1\}@llnl.gov.
    \IEEEcompsocthanksitem Marc Snir is with the Department of Computer Science, University of Illinois Urbana-Champaign. E-mail: snir@illinois.edu.}%
}

\markboth{}
{Chen \MakeLowercase{\textit{et al.}}: Formal Definitions and Performance Comparison of Consistency Models for Parallel File Systems}


\IEEEtitleabstractindextext{
\begin{abstract}
The semantics of HPC storage systems are defined by the consistency models to which they abide.
Storage consistency models have been less studied than their counterparts in memory systems, with the exception of the POSIX standard and its strict consistency model. The use of POSIX consistency imposes a performance penalty that becomes more significant as the scale of parallel file systems increases and the access time to storage devices, such as node-local solid storage devices, decreases. While some efforts have been made to adopt relaxed storage consistency models, these models are often defined informally and ambiguously as by-products of a particular implementation.
In this work, we establish a connection between memory consistency models and storage consistency models and revisit the key design choices of storage consistency models from a high-level perspective. Further, we propose a formal and unified framework for defining storage consistency models and a layered implementation that can be used to easily evaluate their relative performance for different I/O workloads. Finally, we conduct a comprehensive performance comparison of two relaxed consistency models on a range of commonly-seen parallel I/O workloads, such as checkpoint/restart of scientific applications and random reads of deep learning applications. We demonstrate that for certain I/O scenarios, a weaker consistency model can significantly improve the I/O performance. For instance, in small random reads that typically found in deep learning applications, session consistency achieved an 5x improvement in I/O bandwidth compared to commit consistency, even at small scales.
\end{abstract}

\begin{IEEEkeywords}
Consistency model, storage consistency, parallel file system, parallel I/O
\end{IEEEkeywords}
}

\maketitle

\IEEEdisplaynontitleabstractindextext

%
\IEEEpeerreviewmaketitle

\IEEEraisesectionheading{
\section{Introduction}
\label{sec:introduction}}

\IEEEPARstart{H}{igh} performance computing (HPC) systems host parallel applications composed of hundreds to tens of thousands of tightly-coupled processes that typically run for hours or days.
These large-scale applications that run on supercomputers often read and write large amounts of data, spending a significant fraction of their execution time performing I/O \cite{patel2019revisiting, paul2020understanding}.
However, the I/O subsystem, a core component in HPC systems, has not evolved as fast as other components such as compute and interconnect. I/O is emerging as a major bottleneck for many HPC applications. For example, it is shown that I/O can take as much as $85\%$ of the training time of a large-scale deep learning application~\cite{clairvoyant}, the majority of which is due to the random read requests to a large number of training samples.
MuMMI~\cite{mummi}, as another example, is a multi-scale simulation that models the dynamics of RAS proteins. When recording snapshots at a 0.5 ns interval, MuMMI generates over 400 million files, occupying over 1 PB of disk space for a single run, which poses a significant challenge for I/O latency and bandwidth. In order to reduce the I/O demand, compromises such as reducing snapshot frequencies have to be made. As we move beyond the exascale era, the I/O bottleneck will only be exacerbated.

A major constraint on the performance of parallel file systems (PFSs) is their strict adherence to the POSIX consistency model. A consistency model specifies a contract between a programmer and a system, wherein the system guarantees that if the programmer follows the rules, the shared data will be consistent and the results of reading, writing, or updating will be predictable. The POSIX standard~\cite{POSIX} specifies a strong and straightforward consistency model, which requires all writes be immediately visible to all subsequent reads. While the POSIX consistency model is easy to maintain in a single-node environment, it is expensive to maintain at scale. Nevertheless, most widely deployed PFSs, including Lustre~\cite{Lustre}, GPFS~\cite{GPFS}, and BeeGFS~\cite{BeeGFS}, support POSIX consistency.
The cost of supporting POSIX consistency is becoming increasingly unacceptable due to two key reasons: (1) the rapid growth in the scale of HPC systems, which directly increases the software overhead of maintaining POSIX consistency; (2) the emergence of new storage devices such as solid storage devices (SSDs), which greatly improves I/O latency and bandwidth and makes software overhead more significant.
In recent years, many efforts have been made to develop burst buffer (BB) PFSs~\cite{BurstFS, UnifyFS, echofs, GfarmBB, SymphonyFS} (especially user-level systems) with relaxed consistency models, but these models were typically defined ambiguously and informally as by-products of their PFS implementations. This leads to three major issues: (1) Performance: It is challenging for system developers to evaluate and compare the effectiveness of different consistency models; (2) Correctness: It is difficult for programmers to reason about their program or check the correctness of their code; (3) Portability: A program that runs correctly under a given relaxed consistency model is not guaranteed to run correctly on a different model.

When compared to consistency models of shared memory systems (often referred to as \emph{memory models}), storage consistency models (or \emph{storage models} for short) have received far less attention and have not been systematically studied from a higher-level perspective. Similar terminologies and concepts are repeatedly reinvented, and lessons learned from memory models are often overlooked.

To summarize and motivate this work, here we list fundamental questions that have not been clearly answered. The first two focus on comparisons between storage and memory models. The last three focus on storage models and their performance implications.
\begin{enumerate}
    \item What are the reasons for the lack of attention to storage models compared to memory models?
    \item What are the design choices for storage models, and how do they relate to similar choices for memory models?
    \item How do existing storage models compare and what commonalities exist among them? Can they be defined in a unified and formal manner?
    \item What are the performance implications of a storage model?
    \item What are effective methods to evaluate and compare the performance of different storage models?
\end{enumerate}

This work seeks to answer these fundamental questions and develop a better understanding of storage consistency models by conducting a systematic study. Our work makes the following contributions:
\begin{itemize}
    \item We investigate the contributing factors of the limited attention paid to storage models. We show that recent advances in storage techniques are rapidly changing some of these factors (Section~\ref{sec:memory_models_vs_storage_models}).
    \item We revisit the design choices of storage models and relate them to memory models. We highlight the different considerations between memory systems and storage systems for each design choice (Section~\ref{sec:memory_models_vs_storage_models}).
    \item We propose a formal and unified framework for specifying the most widely-used family of storage models (Section~\ref{sec:unified_framework}).
    \item We study the performance implications of storage models. More importantly, we present a ``layered'' implementation that allows for effective performance comparisons between different storage models (Section~\ref{sec:implementation}).
    \item Finally, we conduct a detailed performance comparison between two storage models using a range of common HPC I/O workloads. The results highlight the significant impact storage models can have on I/O performance (Section~\ref{sec:performance_study}).
\end{itemize}

In this work, we focus our study on storage models in the context of parallel file systems for HPC I/O, but the concepts we develop should be generally applicable to other large-scale storage systems.


\section{Background}
\label{sec:background}

This section describes example consistency models from both memory and storage domains, with the aim of introducing their similarities and differences. To prevent confusion, we use terms \emph{store} and \emph{load} when describing memory models and \emph{write} and \emph{read} when describing storage models.

\subsection{Consistency Model: Strong or Relaxed}

Sequential consistency~\cite{lamport_sc} is one of the most intuitive consistency models. It says that the result of any execution is the same as if the operations of all the processors were executed in some sequential order, and the operations of each individual processor appear in this sequence in the order specified by its program. Sequential consistency is considered a strong consistency model because it guarantees operations of a processor are seen to occur in the same order by all processors. The major drawback is that it hinders optimizations that may result in reordering, e.g., store buffers and out-of-order cores.

Relaxed consistency models (weaker than sequential consistency) allow more optimizations but can be counter-intuitive.
Consider the well-known example shown in Table~\ref{tab:load_after_store_example}, where each process loads the value of the variable ($x$ and $y$) stored by the other process.
Intuitively, there are three possible outcomes: $(r_1, r_2) = (0, 100), (100, 0)$ or $(100, 100)$. Sequential consistency guarantees that any execution of this program will produce one of these three results. In reality, most real hardware also allows $(r_1, r_2) = (0, 0)$. For example, x86 systems from Intel uses a relaxed consistency model (often referred to as total store order~\cite{TSO}) that allows reordering non-conflicting store-load pairs, which violates sequential consistency. With this relaxation, store buffers can be used to buffer the expensive stores, so that loads ($L_{12}$ and $L_{22}$) can bypass the previous stores ($L_{11}$ and $L_{21}$).

\begin{table}[htbp]
    \normalsize
    \centering
    \caption{A load-after-store example. All variables are initially zero.}
    \label{tab:load_after_store_example}
    \begin{tabular}{|p{3.2cm}|p{3.2cm}|}
        \hline
        Process 1: & Process 2:\\
        \hline
        $L_{11}: x = 100;$ & $L_{21}: y = 100;$ \\
        $L_{12}: r_1 = y;$ & $L_{22}: r_2 = x;$ \\
        \hline
    \end{tabular}
\end{table}

The core idea behind the relaxed models is that some constraints imposed by stronger models are not necessary for the targeted program, while relaxing such constraints provides significant performance gains. The drawback, however, is that relaxing consistency semantics will likely reduce portability or programmability.

\subsection{Relaxed Memory Models}

\subsubsection{Weak Ordering}
Weak ordering was defined by Dubois et al.,~\cite{weak_ordering} as follows: In a multiprocessor system, memory accesses are weakly ordered if (1) accesses to global synchronizing variables are strongly ordered, (2) no access to a synchronizing variable is issued by a processor before all previous global data accesses have been globally performed, and (3) no access to global data is issued by a processor before previous accesses to a synchronizing variable has been globally performed.

In essence, a system that follows weak ordering needs to be able to recognize synchronization operations. Concurrent accesses to shared memory can violate sequential consistency.
But if all conflicting memory accesses are properly synchronized, then a weakly ordered system will deliver the same result as a system with sequential consistency. Many high-level languages require that programs be race-free, i.e., that conflicting accesses be synchronized.
Consider the example in Table~\ref{tab:weak_ordering_example}, when all operations are identified as data operations, $y$ will not be guaranteed to return $100$ because processors are free to reorder operations. However, if $L_{12}$ and $L_{21}$ are identified by programmers as synchronizations, then $L_{22}$ is guaranteed to return the latest value of $x$ due to the ordering imposed by the synchronizations.

\begin{table}[h]
    \normalsize
    \centering
    \caption{A weak ordering example. All variables are initially zero.}
    \label{tab:weak_ordering_example}
    \begin{tabular}{|p{3.2cm}|p{3.2cm}|}
        \hline
        Process 1: & Process 2:\\
        \hline
        $L_{11}: x = 100;$  & $L_{21}: while(!flag)\{\};$ \\
        $L_{12}: flag = 1$  & $L_{22}: y = x;$ \\
        \hline
    \end{tabular}
\end{table}

\subsubsection{Release Consistency}

Many synchronization operations occur in pairs.
Release consistency~\cite{release_consistency} utilizes this information by explicitly distinguishing them as \emph{release} and \emph{acquire} operations, with the help from programmers. The release operation instructs the processor to make all previous memory accesses globally visible before the release completes, and the acquire operation instructs the processor not to start subsequent memory accesses before the acquire completes.  
In the example of Table~\ref{tab:weak_ordering_example}, $L_{12}$ is a release operation and $L_{21}$ is an acquire operation.
Release consistency is a further relaxation of weak ordering. It allows systems to have different implementations for release and acquire, which leads to better performance at the cost of the increased burden on programmers.

\subsubsection{Entry Consistency}

A major issue of weak ordering and release consistency is that their synchronization operations impose order on memory operations even if they do not conflict, which may add unnecessary overhead. Consider the example in Table~\ref{tab:entry_consistency_example}, to make sure $y$ in $L_{22}$ returns the store to $x$ in $L_{12}$, under weak ordering or release consistency, $L_{13}$ and $L_{21}$ need to be identified as synchronizations. However, this also prohibits reordering $L_{11}$ and $L_{13}$, i.e., $L_{11}$ must complete before $L_{13}$, which is unnecessary if no other process will ever access $w$.
Entry consistency addresses this issue by requiring each ordinary shared data item to be associated with a \textit{synchronization variable}. When an acquire is done on a synchronization variable, only those data guarded by that synchronization variable are made consistent. For instance, in the case of example in Table~\ref{tab:entry_consistency_example}, we can associate $w$ and $x$ with two different synchronization variables, thus allowing $L_{11}$ to bypass $L_{12}$ and $L_{13}$.

\begin{table}[!h]
    \normalsize
    \centering
    \caption{An entry consistency example. All variables are initially zero.}
    \label{tab:entry_consistency_example}
    \begin{tabular}{|p{3.2cm}|p{3.2cm}|}
        \hline
        Process 1: & Process 2:\\
        \hline
        $L_{11}: w = 100;$  & $L_{21}: while(!flag)\{\};$ \\
        $L_{12}: x = 100;$  & $L_{22}: y = x;$ \\
        $L_{13}: flag = 1$  & \\
        \hline
    \end{tabular}
\end{table}


\subsection{Relaxed Storage Models}
\label{subsec:relaxed_storage_models}

The requirements of POSIX consistency essentially impose sequential consistency.
The fundamental problem behind the performance issues stemming from POSIX consistency is that PFSs are ignorant of the application's synchronization logic and the order of I/O operations of different processes. PFSs must make worse-case assumptions and serialize all potentially conflicting I/O operations to guarantee POSIX consistency. 
Alternatively, programmers can provide information on program synchronizations of conflicting I/O operations to the PFS. With this extra information, PFSs can adopt a weaker consistency model, while guaranteeing the same outcome of POSIX consistency.
Wang et al.,~\cite{wang2020hpdc} have studied many such PFSs and their consistency models. Here, we briefly discuss the most commonly used models. 

\subsubsection{Commit Consistency}

Commit consistency is a relaxed consistency model commonly used by recent BB PFSs such as BSCFS~\cite{BSCFS}, UnifyFS~\cite{UnifyFS}, and SymphonyFS~\cite{SymphonyFS}. In commit consistency, ``commit'' operations are explicitly executed by processes. The commit operation conveys synchronization information. I/O updates performed by a process to a file before a commit become globally visible upon return of the commit operation. 
To maintain portability, PFSs adopting commit consistency may use an existing POSIX call to indicate a commit. For example, in UnifyFS~\cite{UnifyFS}, a commit operation is triggered by a \texttt{fsync} call, which applies to all updates performed by a process on a file since the previous commit. Note that finer commit granularity (e.g., committing byte ranges) is also possible, but may add additional overhead if used in a superfluous way.

\subsubsection{Session Consistency}

Commit consistency guarantees all local writes that precede the commit operation become globally visible after the commit operation. However, in many cases, data written is rarely read back by the same application, and even when this happens, usually only a subset of processes perform the reads. Thus, global visibility is not necessary. Session consistency (also known as close-to-open consistency) addresses this issue by defining a pair of synchronization operations, namely, \texttt{session\_close} and \texttt{session\_open}.
Session consistency guarantees that writes by a process become visible to another process (not all processes) when the modified file is closed by the writing process and subsequently opened by the reading process, with the \texttt{session\_close} happening before the \texttt{session\_open}. The idea of session consistency is very similar to that of release consistency for memory models.

Note that we name the two operations \texttt{session\_open} and \texttt{session\_close}, but most existing systems adopting session consistency such as NFS~\cite{NFSv4} do not provide the separate \texttt{session\_open} and \texttt{session\_close} APIs. Rather, they are implied by POSIX \texttt{open/close} (or \texttt{fopen/fclose}) calls, calls that have additional effects---they apply all updates to a file.

\subsubsection{MPI-IO Consistency}
\label{subsec:mpi_io_consistency}

MPI-IO~\cite{MPI-IO} is a part of the MPI standard that defines both communications (message passing) and I/O operations. As the latest standard~\cite{MPI4} states, MPI-IO provides three levels of consistency: sequential consistency among all accesses using a single file handle, sequential consistency among all accesses using file handles created from a single collective open with atomic mode enabled, and user-imposed consistency among accesses other than the above.

The first two cases are the most common cases, and sequential consistency is guaranteed without extra synchronizations. 
In the last case, sequential consistency can be achieved by using a \emph{sync-barrier-sync} construct that imposes orders on the conflicting I/O accesses. 
Here, \emph{sync} is one of \texttt{MPI\_File\_open, MPI\_File\_close or MPI\_File\_sync} that takes in the file handle and flushes the data (writer) or retrieves the latest data (reader). And \emph{barrier} provides a mechanism that imposes order between the two \emph{syncs}. In most cases, this is achieved using MPI calls. For example, \emph{barrier} can be one of the collective communication calls such as \texttt{MPI\_Barrier} and \texttt{MPI\_Allgather}, or a pair of point-to-point calls such as \texttt{MPI\_Send} plus \texttt{MPI\_Recv}.
However, \emph{barrier} is not limited to MPI calls, it can use any mechanism that properly orders the two \emph{sync} calls.

Even though MPI-IO has information on both I/O and MPI communication, it can not assume the synchronization information (i.e., \emph{barrier}) is always available to the system as it may not use MPI calls. MPI-IO consistency is similar to session consistency, but additional optimizations are possible if ordering is imposed through MPI calls, as those are visible to the MPI library (which includes MPI-IO).

\section{Memory Models vs. Storage Models}
\label{sec:memory_models_vs_storage_models}

In this section, we first investigate why relaxed storage consistency models have not gained enough attention and widespread adoption. Next, we analyze the key design choices for consistency models and compare the different considerations between memory systems and storage systems. Finally, we discuss the primary commonality of existing relaxed storage models, introduce the key concepts that will serve as the foundation for our formal definition of these models.

\subsection{Why Relaxed Storage Models are not Widely Adopted}
\label{subsec:programming_hierarchy}

\subsubsection{Programming Hierarchy}

\begin{figure}[htbp]
    \centering
    \includegraphics[width=\linewidth]{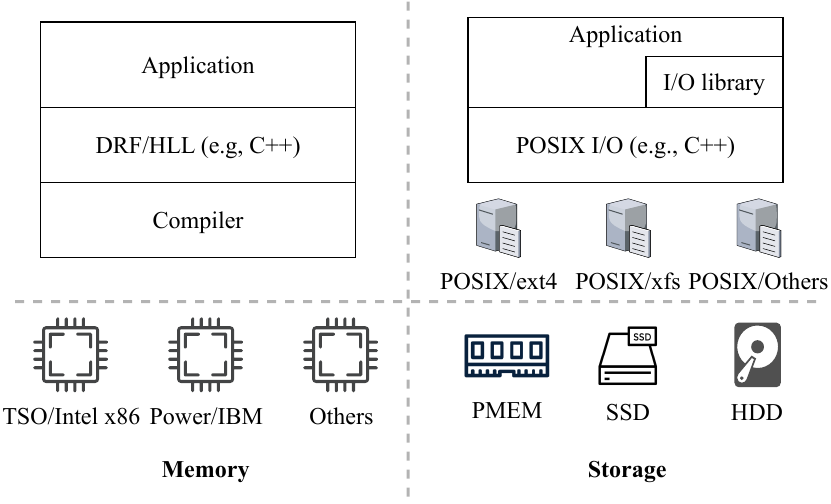}
    \caption{Memory vs. storage programming. The lack of automated support (e.g., a compiler layer) in storage programming hierarchy makes it harder to adopt different consistency models for different hardware.}
    \label{fig:memory_vs_storage_programming}
\end{figure}

The presence of compilers in the memory programming hierarchy (Figure~\ref{fig:memory_vs_storage_programming}(a)) has been an important factor in the adoption of relaxed memory models. Compilers can hide complexity and provide portability, which allows programmers to target a single memory model specified by the high-level programming language (e.g., C++ and Java) without the knowledge of underlying consistency models provided by the CPUs. This way, a suitable consistency model can be selected for the given hardware, without worrying about programmability. For example, C++ allows specifying a different consistency model for each atomic operation; but the semantics of these models is specified by the C++ standard is unrelated to the underlying hardware. 

In contrast, there is no corresponding ``compiler'' layer in the storage programming hierarchy and no low-level hardware-supported consistency model to map onto: Consistency protocols are implemented in software. To achieve programmability and portability, most storage systems choose to implement the same standard, POSIX. Most local file systems (e.g., ext3, ext4, and xtfs) and parallel file systems (e.g., Lustre~\cite{Lustre} and GPFS~\cite{GPFS}) are POSIX-compliant. 

\subsubsection{Software Overhead}

The POSIX consistency semantics prohibit many optimizations. This disadvantage is not so apparent in a single-node system, in which I/O operations are serialized. But maintaining POSIX consistency in an HPC environment can be more costly, as PFSs require distributed locking mechanisms running reliably at large scale to enforce it. Nevertheless, in most scenarios, the software overhead incurred is minor compared to the slow I/O performance of HDDs. Consequently, less attention has been given to alternative consistency models.
However, this is starting to change due to two reasons: the rapid increase in the scale of HPC systems, and the emergence of new, faster storage technologies, such as SSDs. The former directly increases the software overhead required to enforce the same consistency requirements. The latter makes the overhead more significant since I/O operations complete much faster.

\subsection{Design Considerations}

Here we describe several important design considerations of a consistency model and compares memory models (from the perspective of high-level programming languages) with storage models (from the perspective of parallel file systems) for each consideration.

\subsubsection{Synchronization}

Synchronization is critical to a consistency model. It is used to enforce order between potential conflicting accesses. Synchronizations can be performed by explicitly invoking synchronization operations, which is common in both memory models and storage models. Alternatively, synchronizations can be specified in a declarative manner.
For example, high-level languages often provide  keywords (e.g., \texttt{atomic} in C++ and \texttt{volatile} in Java) that can modify ordinary objects to impose extra ordering restrictions on relevant accesses. Such features, however, are less common in storage models.

\subsubsection{Scope of synchronization control}
Both high-level programming languages and parallel file systems provide some limited control on synchronization requirements of executing code. In a high-level language such as C++ this is done at the level of variable declarations (for atomic variables) or atomic access operations (specifying the applicable memory order). The latter is seldom used.
For POSIX file systems, this is done when a file is opened, e.g., with \texttt{O\_SYNC, O\_RSYNC} or \texttt{O\_DSYNC}
flags. Other scopes for such controls are feasible, in both cases. For a programming language, the scope is likely to be a static text scope; for a file system, it is likely to be a file, file range, or an I/O call.

\subsubsection{Atomicity}
Atomicity is often a required property for both memory systems and storage systems. It is key to ensure correctness for applications with conflicting accesses.

\subsubsection{Granularity}
Atomicity is supported in high-level languages with arbitrary granularity. One can specify a primitive object (e.g., \texttt{int}) or a large data structure to be atomic. This granularity needs not to match the granularity of memory operations in hardware; the compiler will implement them using native atomic operations or locks, depending on the granularity. Support for atomic access to larger memory objects will entail additional software overheads.
Similarly, consistency is supported by memory hardware and made visible in high-level languages at the granularity of the smallest accessible datum, namely a byte. But coherence protocols act at the granularity of a cache line (typically, 64 bytes). 
Finer-grain coherence units would require more hardware; coarser-grain coherence units increase the amount of coherence memory traffic and the frequency of false sharing. 

File systems also support storage accesses having arbitrary lengths. POSIX does not guarantee atomicity of reads and writes; the outcome of such operations is well-defined only if conflicting operations are ordered by some means. On the other hand, consistency is maintained at the byte level by POSIX. PFSs' units of coherence are necessarily much coarser, so that fine-grain interleaved accesses by distinct processes can generate a significant amount of coherence traffic and suffer from false sharing.    

\subsubsection{Program text}
In memory systems, compilers see the program text and thus have some information on possible executions of the program. Parallel file systems, on the other hand, have no access to the program text. A PFS is an online system that sees one storage operation at a time.

\subsubsection{Reordering}
The compiler can perform static passes to reorder memory instructions, whereas PFSs are online systems that do not have the ability to make static reorderings. PFSs can perform some limited reorderings by buffering/delaying certain storage operations.

\subsubsection{External information}

When programming on a PFS, as discussed in Section~\ref{subsec:mpi_io_consistency}, programmers sometimes use non-storage operations, e.g., through RPC and message passing, to express their synchronization logic. However, PFSs are generally unaware of synchronization operations. In contrast, memory models are simpler as they assume that all synchronization is done using memory operations.

\subsection{Approach to Formally Define Consistency Models}

The primary commonality of existing relaxed storage models is that they can guarantee sequential consistency for programs that follow certain rules. Those programs share enough information (e.g., the commit calls in commit consistency) with the system so the system can guarantee sequentially consistent execution results even with relaxed storage models.
Such models are said to be in \emph{Sequential Consistency Normal Form (SCNF)}~\cite{adve1993thesis}. SCNF was a term initially defined for memory model formalization, but it applies to storage models as well.
\begin{description}
\item[Sequential Consistency Normal Form:] A consistency model is in sequential consistency normal form iff it guarantees sequential consistency to a set of formally-characterized programs.
\end{description}

The idea of providing sequential consistency semantics to a set of formally-characterized programs was formalized by the data-race-free (DRF) memory models~\cite{adve1990DRF0, adve1993thesis}.
The DRF models exploits the observation that good programming practice dictates that programs be data-race-free; a data race often suggests that there are bugs in the code. The DRF models guarantee sequential consistency for the ``correct'' programs (i.e., without data races) and leave the behavior of the ``incorrect'' programs undefined. 

Unfortunately, unlike the DRF memory model, existing SCNF storage models are typically defined ambiguously and informally as by-products of their PFS implementations. In the next section, we will present a unified and formal framework to specify storage models that are in SCNF. 

\section{A Unified and Formal Framework}
\label{sec:unified_framework}

The SCNF storage models we consider rely on \emph{synchronization information} to achieve sequential consistency.
We call programs that contain adequate synchronization to enforce necessary ordering \emph{properly-synchronized programs}, and the storage models that guarantee sequential consistency to those programs \emph{properly-synchronized SCNF models}. All models we discussed in Section~\ref{subsec:relaxed_storage_models} are properly-synchronized SCNF models.

The formalization of our framework is similar to that of the Java memory model~\cite{java_memory_model} (which adopts the DRF approach but with a much more complex model). Our framework does not make any assumptions about particular synchronization methods; it allows the specific storage model to define its own set of synchronization operations. The key is to define which programs are considered properly-synchronized.

\subsection{Specifying Properly-Synchronized SCNF Models}

We first define two types of storage operations: A storage operation is either a \emph{data storage operation} or a \emph{synchronization storage operation}, defined as follows.

\begin{description}
\item[Data Storage Operations:] These are I/O operations that read or write storage, such as \texttt{fread} or \texttt{fwrite}. Data operations include the specification of the storage location (possibly as a range) to be read or written. Each data operation specifies an object called \emph{synchronization object} that is associated with the requested location, such as a file handle.
\item[Synchronization Storage Operations:] These are I/O operations that may be used to impose an order on data storage operations, such as \texttt{fsync, fopen}, or \texttt{fclose}. Synchronization operations are model-specific, where each synchronization operation includes the specification of a synchronization object.
\end{description}

Further, we consider here the execution of a multiprocess program, in an environment that provides well-defined mechanisms to synchronize concurrent processes, such as MPI message-passing. These mechanisms define a \emph{program order} and \emph{synchronization order} on the executed operations of the program:

\begin{description}
\item[Program Order ($\xrightarrow{po}$):] The program order of a process is a total order on the execution of the process' operations as specified by the program text. To keep the discussion simple, we ignore the extensions needed to deal with multithreaded processes.

\item[Synchronization Order ($\xrightarrow{so}$):] A synchronization order is a partial order specified between operations executed by distinct processes. This partial order is consistent with the program order, and $\xrightarrow{po} \cup \xrightarrow{so}$ is acyclic.
\end{description}

A properly-synchronized SCNF model is then defined as follows. 


\begin{description}
\item[Happens-Before Order ($\xrightarrow{hb}$):] The happens-before order of an execution is the transitive closure of $\xrightarrow{po} \cup \xrightarrow{so}$. The outcome of a parallel execution should be as if all instructions were executed in the order specified by $\xrightarrow{hb}$. Thus, if $ow$ and $or$ are, respectively, a write and a read to the same location, and $ow \xrightarrow{hb} or$, then $or$ will return the value written by $ow$, unless there is another store $ow'$ to the same location such that $ow \xrightarrow{hb} ow' \xrightarrow{hb} or$.

The happens-before order is defined by the semantics of the programming system used. It orders I/O operations executed by the program. It is not necessarily visible to the storage system.


\item[Conflict:] Two data storage operations \textit{conflict} iff their access ranges overlap, and at least one of them is a write.

\item[Minimum Synchronization Construct (\emph{MSC}):] An MSC specifies a minimum sequence of synchronization storage operations required to synchronize two conflicting data operations. An MSC consists of $k$ synchronization storage operations and $k+1$ edges, where $k \geq 0$:
$$
MSC =  \xrightarrow{r_0} S_1 \xrightarrow{r_1} S_2 \xrightarrow{r_2} ... \xrightarrow{r_{k-1}} S_k \xrightarrow{r_k}
$$
For each $i$, $1 \leq i \le k$ and $S_i \in S$, where $S$ is the set of synchronization storage  operations to be defined by the specific consistency model. For each $j$, $0 \leq j \leq k$ and $\xrightarrow{r_j} \in \{\xrightarrow{po}, \xrightarrow{hb}\}$. Note here the choice of $r_j$ can not be trimmed down to just $\xrightarrow{hb}$ as some consistency models may require a synchronization operation of the MSC to be called by one of the conflicting processes, where $r_j = \xrightarrow{po}$.

\item[Properly-Synchronized Relation ($\xrightarrow{ps}$):] Two conflicting data storage operations $X$ and $Y$ are properly synchronized, i.e., $X \xrightarrow{ps} Y$, iff one of the following holds:
\begin{enumerate}
    \item $X$ is a read operation and $X \xrightarrow{hb} Y$.
    \item $X$ is a write operation, and there exists an MSC between $X$ and $Y$ in the happens-before order.
\end{enumerate} 

\item[Storage Race:] Two data storage operations $X$ and $Y$ in an execution form a \emph{storage race} iff they conflict and they are not properly synchronized.

\item[Properly-Synchronized Program:] A program is properly synchronized iff for every sequentially consistent execution of the program, all storage operations can be distinguished by the system as either data or synchronization, and there are no storage races in the execution.

\item[Properly-Synchronized SCNF System:] A system is said to be a properly-synchronized SCNF system iff the result of every run of a properly-synchronized program on the system is the result of a sequentially consistent execution of the program.

\end{description}

Intuitively speaking, the key to achieving sequential consistency is to make sure the program is \emph{storage race free} (i.e., there are no conflicts or conflicts are properly synchronized). Storage race-freedom may require the use of storage synchronization operations, in addition to the synchronization constructs of the parallel programming system. The properly-synchronized SCNF model specifies a set \emph{S} of storage synchronization operations and minimum synchronization constructs (\emph{MSC}) to properly synchronize conflicting I/O operations. 

\subsection{Describing Existing Models}

Our framework provides a formal, but simple, way to capture the specification of properly-synchronized SCNF models, where only \emph{S}  and \emph{MSC} need to be specified for a complete definition.
Table~\ref{tab:example_of_properly_synchronized_scnf_models} demonstrates how to describe the storage models discussed earlier (Section~\ref{subsec:relaxed_storage_models}) using our framework.

\subsubsection{POSIX Consistency}
POSIX consistency can be considered as a special properly-synchronized SCNF model. With POSIX consistency, every write is immediately visible to all subsequent reads without synchronization operations. Here and in the rest of this section, ``subsequent'' is defined according to the happens-before order. Therefore, POSIX consistency has an empty set \emph{S} and an \emph{MSC} of $\xrightarrow{hb}$.

\subsubsection{Commit Consistency}

For commit consistency, there is one synchronization operation, \texttt{commit}. A write to file $f$ becomes visible to all subsequent reads from $f$ upon the return of a subsequent the commit call. 
Most commit-based systems require that the commit is called by the process that performs the writes, by having an \emph{MSC} of $\xrightarrow{po}$ \texttt{commit} $\xrightarrow{hb}$. A relaxed version may allow a process to commit for the updates of other processes, resulting an \emph{MSC} of $\xrightarrow{hb}$ \texttt{commit} $\xrightarrow{hb}$.

\begin{table*}[htb]
    \centering
    \normalsize
    \caption{Specifying properly-synchronized SCNF models using our framework.}
    \label{tab:example_of_properly_synchronized_scnf_models}
    \begin{tabular}{|l|p{5.8cm}|p{7.2cm}|}
        \hline
        Consistency Models & \emph{S} & \emph{MSC}\\
        \hline
        POSIX Consistency & $\{\}$ & $\xrightarrow{hb}$ \\ \hline
        Commit Consistency &  $\{\texttt{commit}\}$ & $\xrightarrow{hb}\texttt{commit}\xrightarrow{hb}$ \\ \hline
        Session Consistency & $\{\texttt{session\_close, session\_open}\}$ & $\xrightarrow{po}\texttt{session\_close}\xrightarrow{hb}\texttt{session\_open}\xrightarrow{po}$ \\ \hline
        MPI-IO Consistency & $\{\texttt{MPI\_File\_sync, }$ & $\xrightarrow{po} s_1 \xrightarrow{hb} s_2 \xrightarrow{po}$ \\ 
        & \texttt{MPI\_File\_close, } & $s_1 \in \{\texttt{MPI\_File\_close, MPI\_File\_sync}\}$ \\
        & \texttt{MPI\_File\_open}$\}$ & $s_2 \in \{\texttt{MPI\_File\_sync, MPI\_File\_open}\}$ \\
        \hline
    \end{tabular}
    \end{table*}
\subsubsection{Session Consistency}
Session consistency specifies two special synchronization operations, $\text{\emph{S}} = \{\text{\texttt{session\_close, session\_open}}\}$. For a write to become visible to a subsequent read, a close-to-open pair has to be performed in between, thus, $MSC = \xrightarrow{po}\texttt{session\_close}\xrightarrow{hb}\texttt{session\_open}\xrightarrow{po}$. The $\xrightarrow{po}$ at the beginning indicates that the \texttt{session\_close} operation has to be performed by the writing process. Similarly, the $\xrightarrow{po}$ at the end indicates that the \texttt{session\_open} operation must be performed by the reading process. Finally, the $\xrightarrow{hb}$ enforces that the \texttt{session\_close} happens before the \texttt{session\_open}.

\subsubsection{MPI-IO Consistency}

As discussed in Section~\ref{subsec:mpi_io_consistency}, MPI-IO provides three levels of consistency. For the first two cases, MPI-IO guarantees sequential consistency without requiring extra synchronizations (just like POSIX consistency). 
In Table~\ref{tab:example_of_properly_synchronized_scnf_models}, we show how to specify the MPI-IO consistency model for the third case. In this case, \texttt{MPI\_File\_close} synchronizes with all subsequent \texttt{MPI\_File\_open} and \texttt{MPI\_File\_sync}. \texttt{MPI\_File\_sync} synchronizes with all subsequent \texttt{MPI\_File\_sync} and \texttt{MPI\_File\_open}. Therefore, there are four possible \emph{MSCs} that can be used to properly synchronize the conflicting accesses:
\begin{itemize}
    \item $\xrightarrow{po}$ \texttt{MPI\_File\_close} $\xrightarrow{hb}$ \texttt{MPI\_File\_open} $\xrightarrow{po}$
    \item $\xrightarrow{po}$ \texttt{MPI\_File\_close} $\xrightarrow{hb}$ \texttt{MPI\_File\_sync} $\xrightarrow{po}$
    \item $\xrightarrow{po}$ \texttt{MPI\_File\_sync} $\xrightarrow{hb}$ \texttt{MPI\_File\_sync} $\xrightarrow{po}$
    \item $\xrightarrow{po}$ \texttt{MPI\_File\_sync} $\xrightarrow{hb}$ \texttt{MPI\_File\_open} $\xrightarrow{po}$
\end{itemize}

In each MSC, the $\xrightarrow{hb}$ imposes the order between the two synchronization operations, and the $\xrightarrow{po}$ enforces that the synchronization operations must be called by the conflicting processes.

\section{An Implementation for Properly-Synchronized SCNF Systems}
\label{sec:implementation}

Now that we have formally defined properly-synchronized SCNF models, the next question is: when should we use a particular consistency model? Another question that immediately follows is: what is the performance difference? Alternatively and more simply, how much performance can we gain from using a weaker consistency model? The answers to these questions are important for both storage system developers and application programmers because they provide information to aid in understanding the trade-off between extra programming effort and extra performance. This information helps system developers choose which consistency models to support and helps application programmers decide whether to port their codes to a storage system with weaker consistency.

\begin{figure}[htbp]
    \centering
    \includegraphics[width=0.95\linewidth]{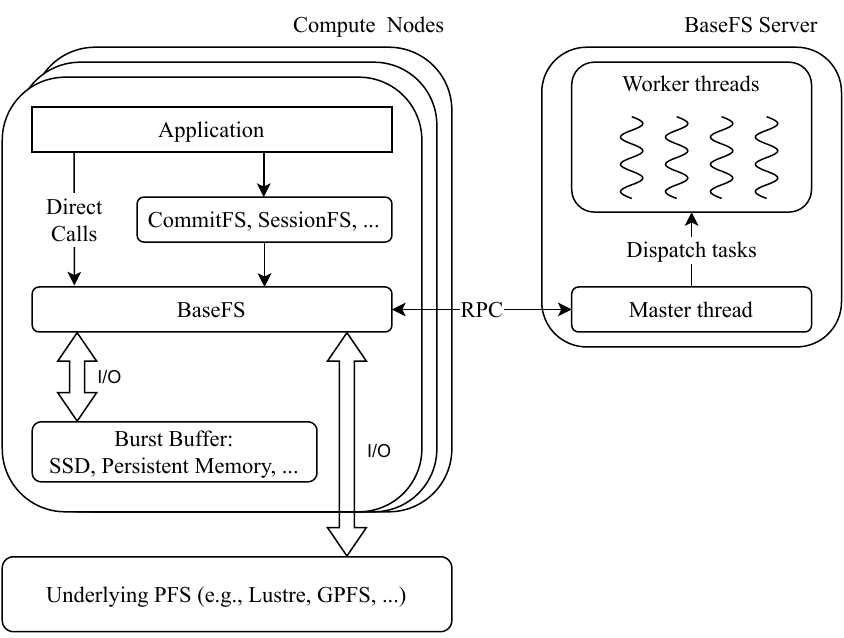}
    \caption{Overview of a layered approach for implementing PFSs with different consistency models.}
    \label{fig:basefs_architecture}
\end{figure}

To answer these questions, we need to conduct a comprehensive performance comparison between different properly-synchronized SCNF models, which requires evaluating PFSs that use those models. 
However, existing PFSs that adopt different consistency models also differ greatly in their implementations and optimizations. \emph{It is difficult to isolate the effect of a consistency model and ever harder to conduct a fair comparison between different consistency models.}
To address this, we present a ``layered'' implementation that allows for an easy performance comparison of different consistency models by keeping, as much as possible, everything other than the consistency model same. An overview of our approach is depicted in Figure~\ref{fig:basefs_architecture}.
We design and implement a ``base-layer'' PFS, called BaseFS, which runs on top of a system-level PFS such as GPFS or Lustre. BaseFS supports the basic functionalities of a PFS with essentially zero optimization. BaseFS buffers reads and writes using burst buffer devices, and flushes data to the underlying PFS only when explicitly instructed. BaseFS provides a very minimum consistency guarantee, but it exposes a set of flexible primitives that can be used to implement custom consistency models.
On top of BaseFS, we can implement PFSs providing different consistency models using these primitives. Since these PFSs use the set of primitives and thus the same underlying implementation, we can limit the impact of other components of the PFS to a very low level. Comparing the performance of these PFSs thus can give us a good understanding of the impact of different consistency models.

In this section we describe BaseFS and two example PFSs, CommitFS and SessionFS, each adopting a different consistency model as suggested by its name.

\subsection{BaseFS}

BaseFS is not designed to be a full-fledged file system.
Our focus is to evaluate the performance implications of different consistency models. As a result, we consider detailed implementation choices, e.g, how to resolve a path and map it to the inode server and how to retrieve file locations given an inode as control variables in our experiments, and we need to make sure that they do not compromise the comparison when evaluating different consistency models.

\subsubsection{Primitives}

Modern PFSs~\cite{GPFS, Lustre, BeeGFS} normally use some kind of locking mechanism to provide sequential consistency. But the lock-based design does not take advantage of the extra information available to the weaker models, like commit consistency and session consistency. Thus, instead of using locking for our BaseFS implementation, we developed a set of flexible primitives (Table~\ref{tab:BaseFS_primitives}) which are more suitable for implementing properly-synchronized SCNF models. 

The BaseFS file system does not provide any implicit guarantee of consistency. Consistency must be enforced by explicit synchronization calls.  The system may store multiple, possibly inconsistent, copies of parts of a file on client nodes, in addition to a (partial) copy on a storage server. 
In BaseFS, the write (\texttt{bfs\_write}) writes to the local copy of the file at the calling client. The read (\texttt{bfs\_read}) is implemented as a \emph{read\_from}: The \emph{owner} argument specifies the client process that will source the data read. The \emph{owner} argument can be retrieved using the \texttt{bfs\_query} call. The read will return the values most recently written by the owner client.

The two key synchronization primitives are \texttt{bfs\_attach} and \texttt{bfs\_query}. The attach call specifies a file range and the issuing client becomes the exclusive owner of those addresses in this range. One can attach only locations that were written by the local process and not flushed.
Essentially, the attach call makes the local writes visible to other processes. It does not guarantee the global visibility of future writes to the same range. Whenever an update needs to be made visible to other processes, an attach call is required.  An attach is not needed if the written data will not be read by other processes.

The query call specifies a file range and returns the current owners of the range. The result is returned in a list of \emph{intervals}. Each interval contains a disjoint subrange and the last attached owner process of that subrange. A query is required to retrieve the the latest attached writes from other processes. In most HPC I/O workloads, this is rare. Typically a process reads from its own writes or from a preexisting file. As a result, the fewer conflicting storage accesses occur, the fewer attach and query calls are needed and thus the lower is the overhead.

\begin{table*}[htbp]
\centering
\caption{The most relevant primitives of BaseFS}
\label{tab:BaseFS_primitives}
\begin{tabular}{|p{\linewidth}|}
    \hline
    $\bullet$ \texttt{bfs\_file\_t* bfs\_open(const char* pathname)} \\
    \tb{Description:} Opens the file whose pathname is the string pointed to by \textit{pathname}, and associates a BaseFS file handle (\texttt{bfs\_file\_t}) with it. This file handle is an opaque object and can be used by subsequent I/O functions to refer to that file. The file is always opened in read-write mode. Append mode is not supported. The file offset used to mark the current position within the file is set to the beginning of the file.\\
    \tb{Return Value:} Upon successful completion, the function returns a pointer to the BaseFS file handle; otherwise, a NULL pointer is returned.\\ \hline
    $\bullet$ \texttt{int bfs\_close(bfs\_file\_t* file)} \\
    \tb{Description:} Causes the file handle pointed to by \textit{file} to be released and the associated file to be closed. Any buffered data is discarded (not flushed as in in POSIX). Whether or not the call succeeds, the file handle is disassociated from the file. \\
    \tb{Return Value:} Upon successful completion, the function returns 0; otherwise, it returns -1. \\
    \hline
    $\bullet$ \texttt{ssize\_t bfs\_write(bfs\_file\_t* file, const void* buf, size\_t size)} \\
    \tb{Description:} Writes \textit{size} bytes of data from the buffer pointed by \textit{buf} to the specified \textit{file}. 
    The file-position indicator of the calling process is advanced by the number of bytes successfully written. 
    The write becomes immediately visible to the writing process, but it is not guaranteed to be visible to other processes after the call. \\ 
    \tb{Return Value:} Upon successful completion, the function returns the number of bytes written; otherwise, it returns -1.\\ \hline
    $\bullet$ \texttt{ssize\_t bfs\_read(bfs\_file\_t* tf, void* buf, size\_t size, bfs\_addr\_t* owner)} \\
    \tb{Description:} Reads \textit{size} bytes of data from the specified \textit{file} to the buffer pointed to by \textit{buf}. 
    The file-position indicator of the calling process is advanced by the number of bytes successfully read.
    This function returns the most up-to-date buffered write of the specified \textit{owner} process. The function will fail if the \textit{owner} process does not own the specified range. If \textit{owner} is \texttt{NULL}, the function will directly read from the underlying PFS. \\
    \tb{Return Value:} Upon successful completion, the function shall return the number of bytes successfully read; otherwise, it returns -1. \\
    \hline
        
$\bullet$ \texttt{int bfs\_attach(bfs\_file\_t* file, size\_t offset, size\_t size)} \newline
        \tb{Description:} Attaches the range from \textit{offset} to \textit{offset+size-1} in \textit{file} to the calling process.  This function makes the most recent buffered writes of the calling process to the specified range visible and available to all processes. Overlapping ranges that were attached by other processes shall be overwritten. The data covered by the specified range must have been written locally. It is allowed to attach partially a previous write, but attaching unwritten bytes is erroneous.
        \newline
        \tb{Return Value:} Upon successful completion, 0 is returned. Otherwise, -1 is returned. \\
        \hline
        
        $\bullet$ \texttt{int bfs\_attach\_file(bfs\_file\_t* file)} \newline
        \tb{Description:} Attaches all locally buffered writes by the calling process to \textit{file}. Overlapping ranges that were attached by other processes shall be overwritten. The function is a no-op if no buffered writes exist. 
        \newline
        \tb{Return Value:} Upon successful completion, 0 is returned. Otherwise, -1 is returned. \\
        \hline
        $\bullet$ \texttt{int bfs\_query(bfs\_file\_t* file, size\_t offset, size\_t size, bfs\_interval\_t** intervals, \newline int* num\_intervals)} \newline
        \tb{Description:} Returns the attached subranges of \textit{file} included in the range of [\textit{offset, offset+size-1}].
        The result is written to \textit{intervals} and \textit{num\_intervals}, where \textit{intervals} contains a list of file ranges and the owner process of each range.\newline
        \tb{Return Value:} Upon successful completion, 0 is returned. Otherwise, -1 is returned. \\
        \hline
        
        $\bullet$ \texttt{int bfs\_query\_file(bfs\_file\_t* file, bfs\_interval\_t** intervals, int* num\_intervals)} \newline
        \tb{Description:} Returns all attached ranges of \textit{file}.
        The result is written to \textit{intervals} and \textit{num\_intervals}, where \textit{intervals} contains a list of file ranges and the attached process of each range. \newline
        \tb{Return Value:} Upon successful completion, 0 is returned. Otherwise, -1 is returned. \\
        \hline
        
        $\bullet$ \texttt{int bfs\_detach(bfs\_file\_t* file, size\_t offset, size\_t size) } \newline
        \tb{Description:} Detaches currently attached ranges in \textit{file} that overlap with range of [\textit{offset, offset+size-1}] of the \textit{file}. The function removes the specified range from the local buffer, and makes the buffered writes covered by the range no longer visible to all processes. If the data is needed for later reads, then a \texttt{bfs\_flush} call should be made before detaching. The function fails if the specified range was not attached before.
        \newline
        \tb{Return Value:} Upon successful completion, 0 is returned. Otherwise, -1 is returned. \\
        \hline
        
        $\bullet$ \texttt{int bfs\_detach\_file(bfs\_file\_t* file)} \newline
        \tb{Description:} Detaches all ranges of \textit{file} that are currently attached to the calling process. The function is a no-op if no attached ranges exist.\newline
        \tb{Return Value:} Upon successful completion, 0 is returned. Otherwise, -1 is returned. \\
        \hline
        
        $\bullet$ \texttt{int bfs\_flush(bfs\_file\_t* file, size\_t offset, size\_t size) } \\
        \tb{Description:} Flushes the locally buffered data in the range from \textit{offset} to \textit{offset+size-1} of \textit{file} to the underlying PFS. Previously attached updates of the same range will remain available to all processes until the detach call.\\
        \tb{Return Value:} Upon successful completion, the function returns 0; otherwise, it returns -1.\\
        \hline

        $\bullet$ \texttt{int bfs\_flush\_file(bfs\_file\_t* file)} \\
        \tb{Description:} Flushes all the locally buffered data (if any) of \textit{file}. The function is a no-op if no locally buffered data exists. \\
        \tb{Return Value:} Upon successful completion, the function returns 0; otherwise, it returns -1.\\
        \hline
        $\bullet$ \texttt{ssize\_t bfs\_seek(bfs\_file\_t* tf, size\_t offset, int whence);} \\
    \tb{Description:} Sets the file-position indicator for \textit{file}. The new position, measured in bytes from the beginning of the file, is obtained by adding \textit{offset} to the position specified by \textit{whence}. The specified point is the beginning of the file for SEEK\_SET, the current value of the file-position indicator for SEEK\_CUR, or end-of-file (EOF) for SEEK\_END.
    Reads from never written locations before the EOF are filled with zeros. Reads from locations beyond the EOF return undefined values.
    The function by itself is not changing the end-of-file location. \\
    \tb{Return Value:} Upon successful completion, the function returns the current file-position indicator; otherwise, it returns -1.\\
    \hline

    $\bullet$ \texttt{ssize\_t bfs\_tell(bfs\_file\_t* file);} \\
    \tb{Description:} This function obtains the current value of the file-position indicator for  \textit{file}.\\
    \tb{Return Value:} Upon successful completion, the function returns the current value of the file-position indicator for the file handle measured in bytes from the beginning of the file. Otherwise, it returns -1. \\
    \hline
        
$\bullet$ \texttt{int bfs\_stat(bfs\_file\_t* file, struct stat* buf)} \\
    \tb{Description:} This function  obtains information about  \textit{file}, and writes it to the area pointed to by \textit{buf}. Currently, BaseFS only maintains the file size attribute (i.e., \texttt{st\_size} of \texttt{struct stat}), all other attributes are ignored. \\
    \tb{Return Value:} Upon successful completion, 0 is returned. Otherwise, -1 is returned. \\
    \hline  
    \end{tabular}
\end{table*}

\subsubsection{Implementation}

Again, the top priority of BaseFS is not to achieve the best performance, but to enable effective comparisons between different consistency models. Therefore, our implementation is fairly straightforward, without complicated optimizations such as distributed servers and namespace partitioning. These optimizations will be equally beneficial to the PFSs built on top of BaseFS (e.g., CommitFS and SessionFS), and would not add additional value to the comparison.

As shown in Figure~\ref{fig:basefs_architecture}, BaseFS is implemented as a user-level BB file system with a focus on data operations. Reads and writes are directly fulfilled by the BB devices without any memory caching. A limited number of metadata operations (e.g., \texttt{stat}) and attributes (e.g., EOF) are supported. 
In BaseFS, each client process buffers its writes (\texttt{bfs\_write}) using node-local BB devices.  We assume that the BB devices are large enough to accommodate the entire storage required for a job execution (no system-initiated flushes). At a read call (\texttt{bfs\_read}), the client reads from the buffer of the specified owner (which can be itself). If the requested range is not owned by any client, the client reads from the underlying PFS to obtain the latest flushed data.

We use a single global server to handle messages from clients. These messages are generated only by the synchronization primitives, the write and read primitives do not involve the global server.
The global server is multithreaded where the master thread handles all communications and the remaining threads run an identical worker routine. Each worker maintains a FIFO queue that holds client requests. When a new client request (e.g., a query request) is received, the master thread creates a new task and appends it to one worker's task queue. The worker is selected in a round-robin manner.
Once the task is completed by the specified worker, the server will send back the result to the requesting client.
Next, we go through the tasks triggered by the synchronization primitives:
\begin{itemize}
    \item \tb{Attaching}: When a client process invokes a \texttt{bfs\_attach*} primitive, it notifies the server that it will be responsible for reads from the specified file range. In other words, the client declares itself as the owner of the most recent update to the specified range. The ownership is exclusive, the caller of \texttt{bfs\_attach*} will take over the ownership in the case when the same range has been previously attached by another process. 
    The subsequent queries (\texttt{bfs\_query}) to the same range will return an exclusive owner.
    Other clients can later use \texttt{bfs\_read} to directly fetch the data from the owner's buffer without going through the underlying PFS. 
    \item \tb{Detaching}: A client detaches from a previously attached file range to relinquish ownership. After detaching, the owner does not own the range anymore and it will not be responsible for future \texttt{bfs\_read} calls to the detached range. If the data needs to be preserved for future reads, then a \texttt{bfs\_flush} call is required before detaching.
    \item \tb{Querying}: A client issues a \texttt{bfs\_query} call to ask the server who owns the most up-to-date data of the given range, i.e., who performed the last attach to the same range. The server will respond with a list of sub-ranges (since the queried range may cover multiple attach operations) along with their owners' information. An empty list will be returned if no one has attached locations in the range yet.
\end{itemize}

The global server maintains a per-file interval tree (noted as \emph{global interval tree}) to keep track of the attached file ranges. Internally, BaseFS uses an augmented self-balancing binary search tree to implement this interval tree. Each interval (or each node of the tree) has the form of $\langle O_s, O_e, Owner \rangle$, where $O_s$ and $O_e$ are the start and end offset of a file range, and $Owner$ stores the information of the most recent client who attached the range. Note that the interval tree keeps only the most recent attach and does not store any histories. A new interval is inserted upon each attach request. At the insertion time, the server checks the existing intervals to decide if they need to be split or deleted. An existing interval is split if it partially overlaps with the new interval and has a different owner; it is deleted if it is fully contained in the new interval. The server also merges intervals belonging to the same client with contiguous ranges. This reduces the number of intervals and accelerates future queries. 
When the server receives a detach request, it consults the interval tree and checks whether the same client still owns the entire range. It is possible that other clients has overwritten the same range and became new owners. In that case, the detach will simply be a no-op. Otherwise, the detach request succeeds (with possible splits), and the interval is removed from the tree.

Each client process also maintains a similar interval tree (noted as \emph{local interval tree}) for each file. It is used to keep track of locally written ranges and their mappings to the local burst buffer files. Specifically, each interval of the local interval tree has the form of $\langle O_s, O_e, B_s, B_e, attached \rangle$, where $O_s$ and $O_e$ indicate the range of a write to the targeted PFS file, $B_s$ and $B_e$ indicate where the range is buffered on the local burst buffer file, and $attached$ indicates whether the write has been attached or not. At each write (\texttt{bfs\_write}), a new interval will be inserted into the local interval tree. There will be no split because all writes are from the same client. Contiguous intervals are merged as in the global interval tree.
The \texttt{bfs\_attach} primitive is used to attach the writes to one contiguous file range, while  
the \texttt{bfs\_attach\_file} primitive attaches all local writes to the file. Both calls will pack and send all supplied information using a single RPC request. Moreover, both calls will check the local interval tree to make sure the same range is not attached twice, and the attached ranges were previously written by the local process.

As mentioned above, a client can respond to read requests from other clients after an attach call. This client-to-client data transfer can be performed efficiently using RDMA. For this to work, each client process needs to spawn a separate thread to listen to the incoming \texttt{bfs\_read} requests. This increases CPU usage but can significantly improve read performance, assuming RDMA is faster than disk I/O (i.e., reading directly from the underlying PFS). 

\subsection{CommitFS and SessionFS}

With BaseFS, we can easily implement a PosixFS, CommitFS, and SessionFS on top. Table~\ref{tab:scnf_model_implementation} shows the APIs exposed by each along with their internal implementations using the BaseFS primitives. The primary difference in their implementations is the placement of attach and query primitives. The stronger the model, the more frequently the attach and query primitives are needed. For example, to achieve POSIX consistency, an attach call has to be invoked by each write, and a query call has to be invoked by each read. In comparison, CommitFS only performs attach at the commit time, though query is still needed ahead of every read operation. 

As for SessionFS, a query is performed at the session open time, and an attach is performed at the session close time. Within a session, multiple write and read calls can be executed without any query or attach.

\begin{table*}[htbp]
    \centering
    \normalsize
    \caption{CommitFS and SessionFS: the exposed APIs and their implementations. POSIX consistency is included for dis.}
    \label{tab:scnf_model_implementation}
    \begin{tabular}{|l|l|l|l|}
        \hline
        File System & Storage Model & Key API & Implementation \\ \hline
        PosixFS  & POSIX consistency & \texttt{open} & \texttt{bfs\_open} \\
                 & & \texttt{close} & \texttt{bfs\_close} \\
                 & & \texttt{write}  & \texttt{bfs\_write; bfs\_attach} \\
                 & & \texttt{read}   & \texttt{bfs\_query; bfs\_read} \\ \hline
        CommitFS & Commit consistency & \texttt{open} & \texttt{bfs\_open} \\
                 & & \texttt{close} & \texttt{bfs\_close} \\
                 & & \texttt{write}  & \texttt{bfs\_write} \\
                 & & \texttt{read}   & \texttt{bfs\_query; bfs\_read} \\ 
                 & & \texttt{commit} & \texttt{bfs\_attach\_file} \\
                 \hline
        SessionFS & Session consistency & \texttt{open} & \texttt{bfs\_open} \\
                  & & \texttt{close} & \texttt{bfs\_close} \\
                  & & \texttt{write}  & \texttt{bfs\_write} \\ 
                  & & \texttt{read}   & \texttt{bfs\_read} \\ 
                  & & \texttt{session\_open}   & \texttt{bfs\_query\_file} \\
                  & & \texttt{session\_close}  & \texttt{bfs\_attach\_file} \\
                  \hline
    \end{tabular}
\end{table*}

\section{The Impact of Consistency Models on I/O Performance}
\label{sec:performance_study}

This section studies the impact of consistency models on I/O performance.
First, we evaluate the performance of commit consistency and session consistency using benchmarks that represent common HPC I/O patterns. Then we perform two case studies to further understand the performance disparity caused by different consistency models. The first case study is of the I/O behavior of the Scalable Checkpoint/Restart (SCR) library~\cite{SCR}, while the second case study is of the I/O behavior of the training phase of distributed deep learning applications. We note that in all cases, we only consider I/O operations and do not perform any computation or communication.

We performed all experiments on the Catalyst system located at Lawrence Livermore National Laboratory. Catalyst is a Cray CS300 system, where each compute node consists of an Intel Xeon E5-2695 with two sockets and 24 cores in total, with 128GB memory. The nodes are connected via IB QDR. The operating system is TOSS 3. Slurm is used to manage user jobs. The underlying PFS is an LLNL customized version of Lustre, 2.10.6\_2.chaos. Each compute node is equipped with an 800GB Intel 910 Series SSD, which serves as the burst buffer device. The peak sequential write bandwidth of the node-local SSD is 1GB/s, and its peak sequential read bandwidth is 2GB/s. We repeated all runs at least 10 times, and the average results are reported.

\subsection{I/O of Scientific Applications}


From a PFS perspective, within each file, there are three common parallel I/O access patterns: (1) \emph{Contiguous}, where multiple processes access the file in a contiguous manner (normally without gaps); (2) \emph{Strided}, where multiple processes access the file in an interleaved manner (often with a fixed stride); and (3) \emph{Random}, multiple processes access the file in a random manner.
The random access pattern is commonly observed in deep learning applications, where multiple processes randomly load samples to feed the neural network. On the other hand, contiguous and strided access patterns are commonly used in parallel scientific applications for performing logging, checkpointing, and outputting snapshots.

We constructed synthetic workloads to simulate common HPC I/O scenarios. Each workload consists of a write phase and/or a read phase, and the read phase begins only after the write phase is complete. Additionally, all processes operate on a single shared file, resulting in an $N\text{-to-}1$ access pattern, where $N$ is the total number of processes. The access pattern within the shared file for each phase (contiguous, strided, or random) can be determined at runtime. The workload can be run on either commit consistency or session consistency using the corresponding APIs provided by CommitFS or SessionFS. The other aspects of the I/O behavior are controlled by the set of parameters summarized in Table~\ref{tab:workload_parameters}.

\begin{table}[htbp]
    \centering
    \normalsize
    \caption{Parameters of the synthetic I/O workloads.}
    \label{tab:workload_parameters}
    \begin{tabular}{|cp{0.8\linewidth}|}
    \hline
    $n_w$ & Number of writing nodes. All processes of a writing node perform only writes. \\
    $n_r$ & Number of reading nodes. All processes of a reading node perform only reads.\\
    $n$ & Total number of nodes; $n = n_r + n_w$. \\
    $p$ & Number of processes per node. Each node runs an equal number of processes.\\
    $m_w$ & Number of writes performed by each process. Each writing process performs the same number of writes. \\
    $m_r$ & Number of reads performed by each process. Each reading process performs the same number of reads. \\
    $s$ & Access size of each I/O operation. All I/O operations have the same access size.\\
    \hline
    \end{tabular}
\end{table}

We used the four configurations shown in Table~\ref{tab:n_1_workload} to conduct the experiments. Each was run on up to 16 nodes with 12 processes per node. In our experiments, write nodes and read nodes did not overlap, so $n_w$ and $n_r$ always added up to $n$. In all runs, we set $m_w = m_r = 10$. Additionally, to understand the impact of a consistency model on scenarios with different accesses sizes, all experiments were run with two different access sizes: 8KB for small accesses and 8MB for large accesses. The file system was purged before the start of each run.

\begin{table}[htbp]
    \centering
    \normalsize
    \caption{Configurations for evaluating the impact of consistency models on common HPC I/O scenarios.}
    \label{tab:n_1_workload}
    \begin{tabular}{|l|l|l|l|l|}
        \hline
        Code name & Write phase & Read phase & $n_w$ & $n_r$ \\ \hline
        CN-W & Contiguous & N/A & n & 0 \\
        SN-W & Strided & N/A & n & 0 \\
        CC-R & Contiguous & Contiguous & $\frac{n}{2}$ & $\frac{n}{2}$  \\
        CS-R & Contiguous & Strided & $\frac{n}{2}$ & $\frac{n}{2}$  \\
        \hline
    \end{tabular}
\end{table}

\subsubsection{Write-only workloads}
\begin{figure}[htbp]
    \centering
    \begin{subfigure}[b]{\linewidth}
        \centering
        \includegraphics[width=\textwidth]{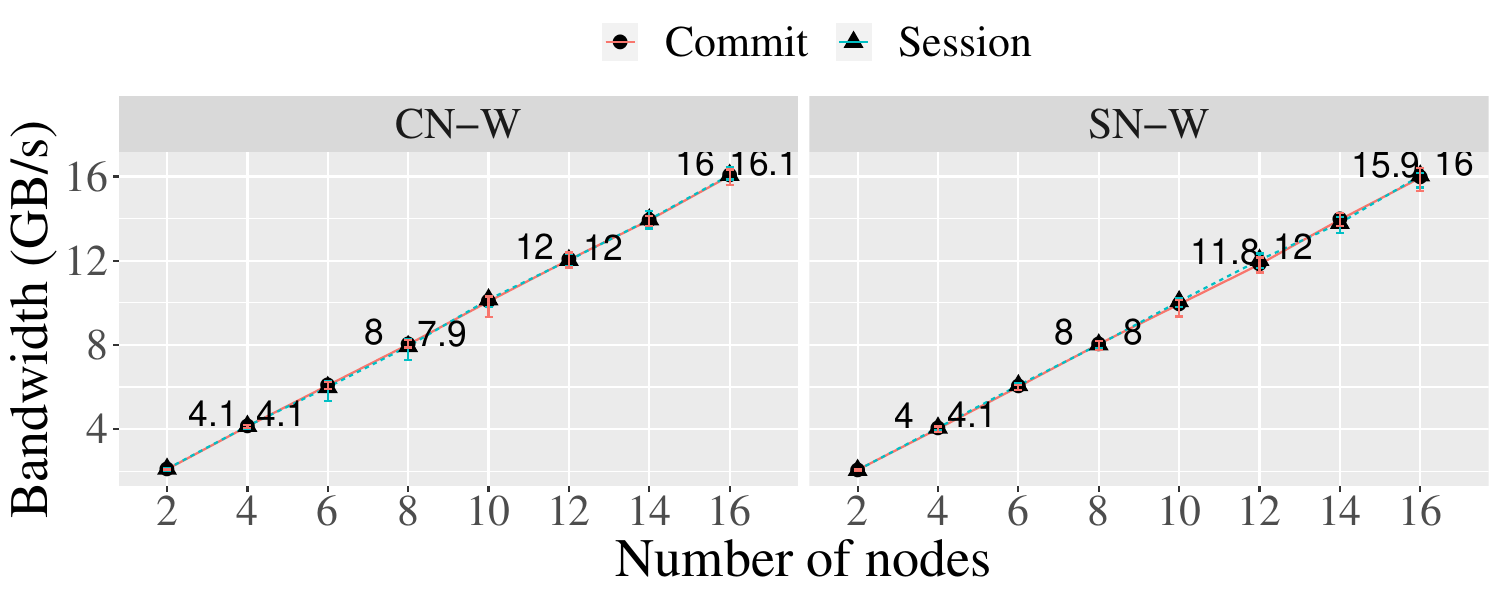}
        \caption{$S = 8$MB}
        \label{fig:iobench_cnw_snw_10x8mb}
    \end{subfigure}
    \begin{subfigure}[b]{\linewidth}
        \centering
        \includegraphics[width=\textwidth]{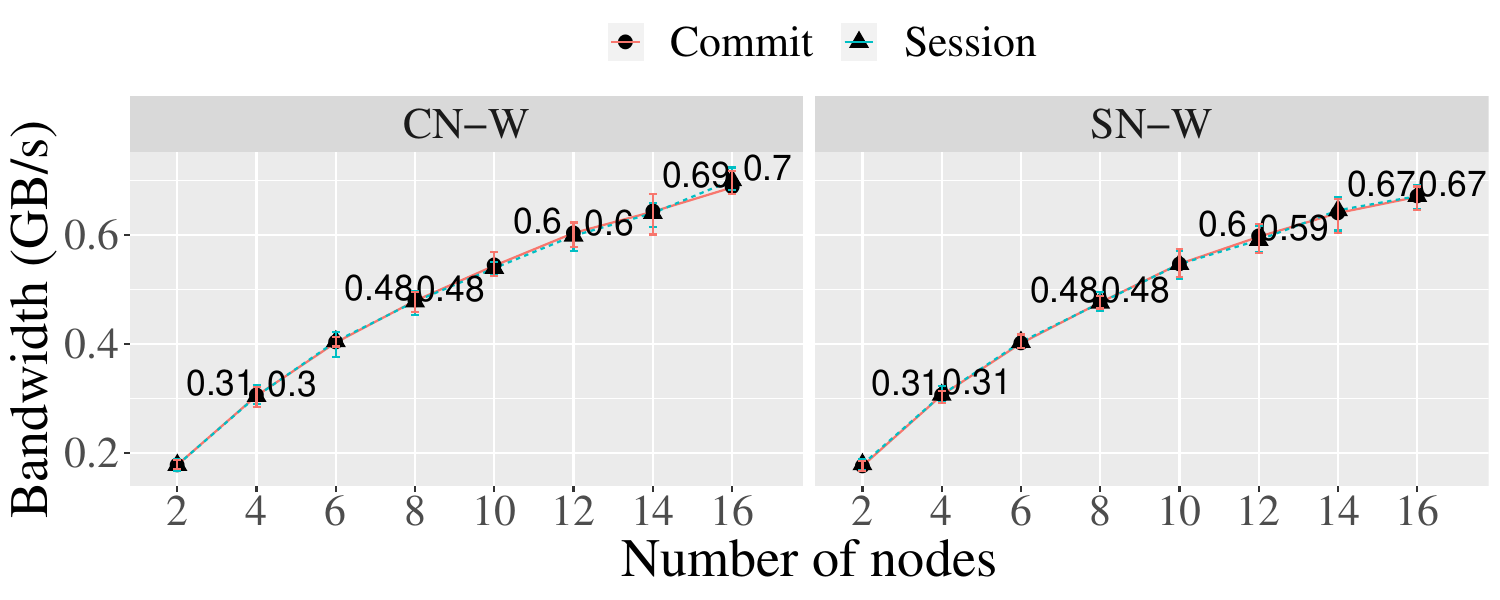}
        \caption{$S = 8$KB}
        \label{fig:iobench_cnw_snw_10x8kb}
    \end{subfigure}
    \caption{Write bandwidth of CN-W and SN-W with 8MB and 8KB access sizes.}
    \label{fig:iobench_cnw_snw}
\end{figure}

The first two configurations, CN-W and SN-W, are write-only and differ only in how writes are performed by the collaborating processes.
Figure~\ref{fig:iobench_cnw_snw} shows their write bandwidths.
With the use of node-local SSDs as burst buffers, all writes are buffered by process-private cache files, which essentially converts the $N\text{-}1$ writes (contiguous or strided) to $N-N$ contiguous writes. Therefore, for both consistency models, the performance of CN-W and SN-W were about the same.

Since the file system is empty when the writes start, \texttt{session\_open} became a no-op, and \texttt{session\_close} performed the same task as \texttt{commit}, thus session consistency and commit consistency achieved similar bandwidths. 

Finally, small writes yielded a worse performance as the small access sizes cannot saturate the bandwidth. When performing large writes, both access patterns were able to achieve the peak write bandwidth, regardless of the consistency model. This is because the overhead required by the consistency model is insignificant compared to the time needed to write to SSD.

\subsubsection{Read-after-write workloads}
The last two configurations, CC-R and CS-R, demonstrate the impact of consistency models on the read bandwidth of workloads with different access patterns. In these configurations, half of the nodes are used for writing and the other half for reading the data back. In CC-R, writes and reads are done contiguously, so each read node reads from only one write node. In contrast, in CS-R, reads are strided, which requires each read node to receive data from multiple write nodes and may cause contention.

\begin{figure}[htbp]
    \centering
    \begin{subfigure}[b]{\linewidth}
        \centering
        \includegraphics[width=\textwidth]{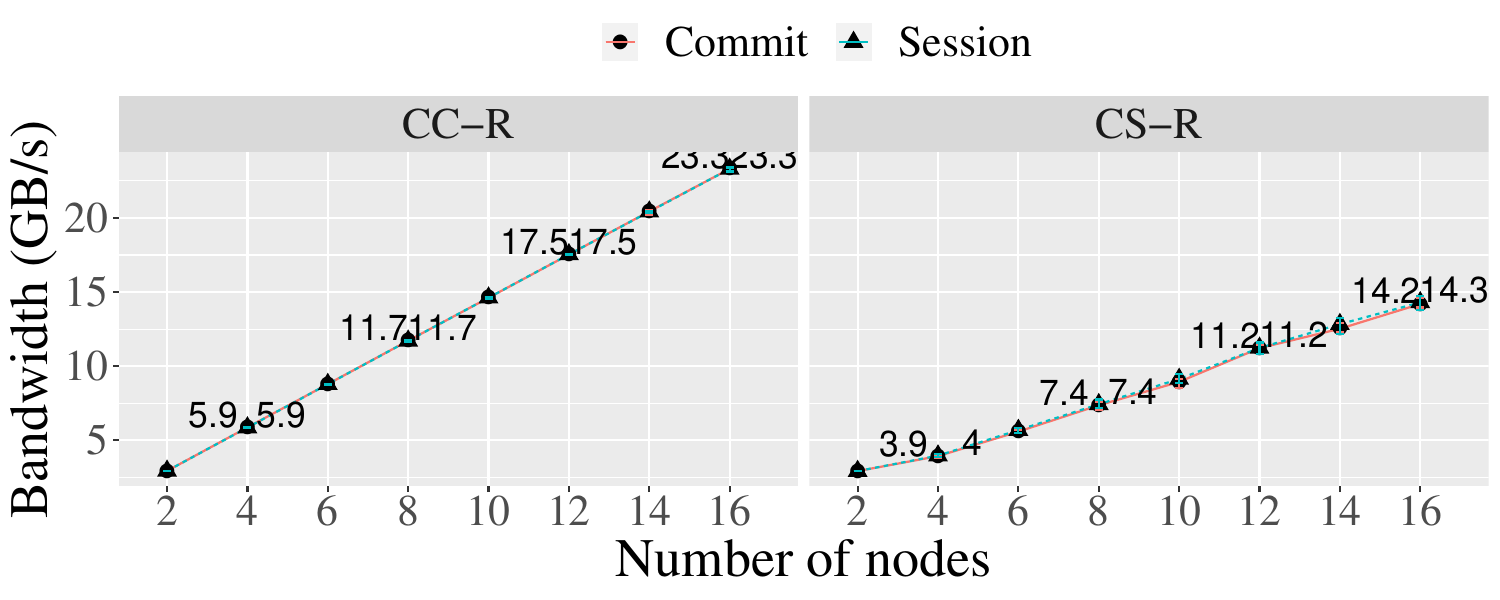}
        \caption{$S = 8$MB}
        \label{fig:iobench_ccr_csr_10x8mb}
    \end{subfigure}
    \begin{subfigure}[b]{\linewidth}
        \centering
        \includegraphics[width=\textwidth]{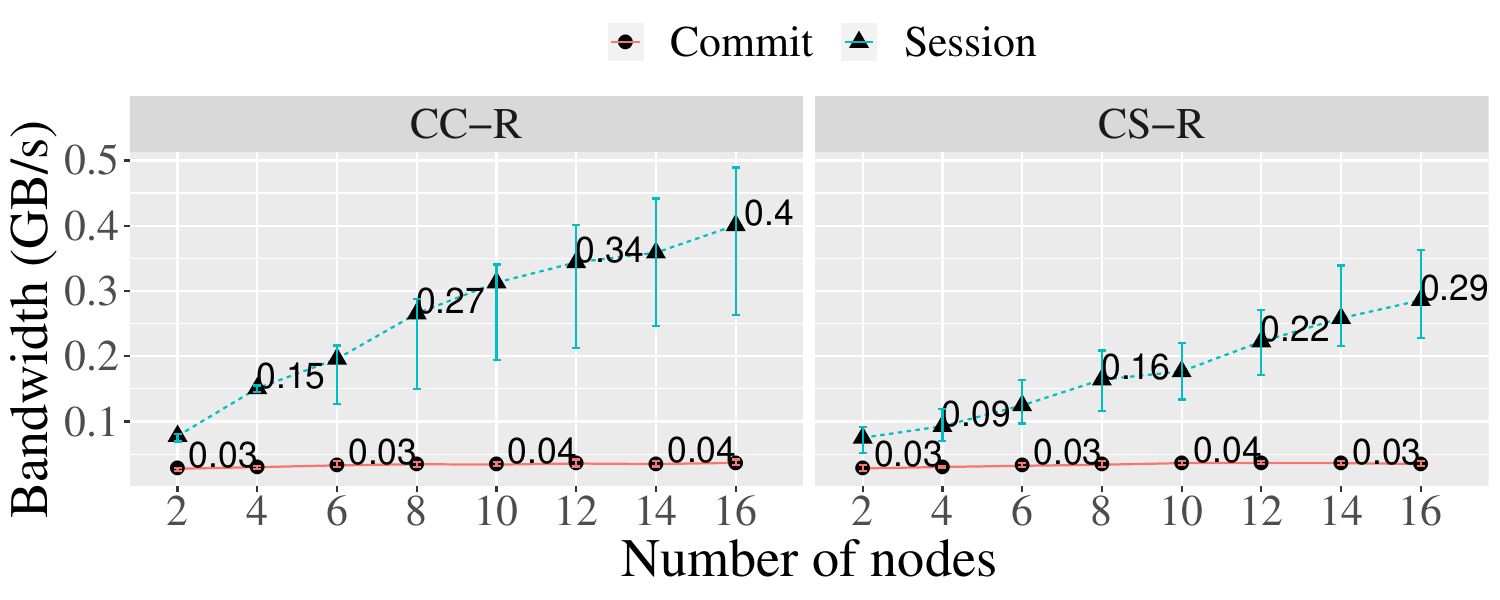}
        \caption{$S = 8$KB}
        \label{fig:iobench_ccr_csr_10x8kb}
    \end{subfigure}
    \caption{Read bandwidth of CC-R and CS-R with 8MB and 8KB access sizes.}
    \label{fig:iobench_ccr_csr}
\end{figure}

The results in Figure~\ref{fig:iobench_ccr_csr} demonstrate that CC-R outperforms CS-R under both consistency models and access sizes. For large reads (Figure~\ref{fig:iobench_ccr_csr_10x8mb}), the impact of consistency models on the bandwidth is negligible.
However, for small reads (Figure~\ref{fig:iobench_ccr_csr_10x8kb}), session consistency achieved better performance and scalability than commit consistency. This is because commit consistency issues an RPC query every time it performs a read, and when the I/O of a read completes quickly, the software overhead becomes the I/O bottleneck, especially when many read requests are performed concurrently. In contrast, session consistency only queries once at the session open time, and the overhead is amortized over a number of reads. 
Lastly, we observed a high variance in the bandwidth of session consistency. To verify whether this was caused by network or system congestion, we repeated the same experiments multiple times at different times of the day and found consistent results. A further investigation (where we used a single node and excluded the communication time) showed that the SSD itself had high variance in small read performance, which we believe is due to normal wear and tear, as SSDs on Catalyst are rather old.  We confirmed this hypothesis by conducting the same experiments on a newer machine (Expanse at San Diego Supercomputer Center), which showed very little variance.

\subsection{Case Study: I/O of Scalable Checkpoint/Restart}

In this subsection, we study the I/O behavior of SCR~\cite{SCR} for checkpointing and restarting HACC-IO~\cite{hacc-io} using an emulator. SCR is a scalable multi-level checkpointing system that supports multiple types of checkpoints with varying costs and levels of resiliency.
The slowest but most resilient level writes to the system-wide PFS, which can withstand an entire system failure. Faster checkpointing for the most common failure modes involves using node-local storage, such as RAM and SSD, and implementing cross-node redundancy schemes.

In our emulation, we consider the most common case where SCR uses node-local storage only. We use the ``Partner'' redundancy scheme, where SCR writes checkpoints to local storage and also copies each checkpoint to storage local to a partner process from another failure group. This scheme requires twice the storage space, but it can withstand failures of multiple devices, so long as a device and the corresponding partner device that holds the copy do not fail simultaneously. To be specific, in our experiment, at the checkpoint phase, SCR buffers the checkpoint data in memory (local and partner) and then flushes it to the SSDs (local and partner) using a file-per-process access pattern. At the restart time, SCR reads directly from the memory buffer assuming the checkpoint data is still accessible.

The client of SCR is HACC-IO, which produces the actual checkpoint data. At each checkpoint step, HACC-IO writes out 9 arrays of the same length, each containing all particle values of a different physical variable. The total data size is determined by the number of particles, which we set to 10 million in our experiment.
Furthermore, the experiment was run with one spare node, and we assumed a single-node failure. When running with $n$ nodes, during the checkpoint phase, $n-1$ nodes wrote to the node-local SSD, with a copy buffered in local memory. During the restart phase, $n-2$ nodes read directly from the local memory buffer, and the spare node receives the checkpoint through MPI from the partner of the failed node.

\begin{figure}[htbp]
    \centering
    \includegraphics[width=\linewidth]{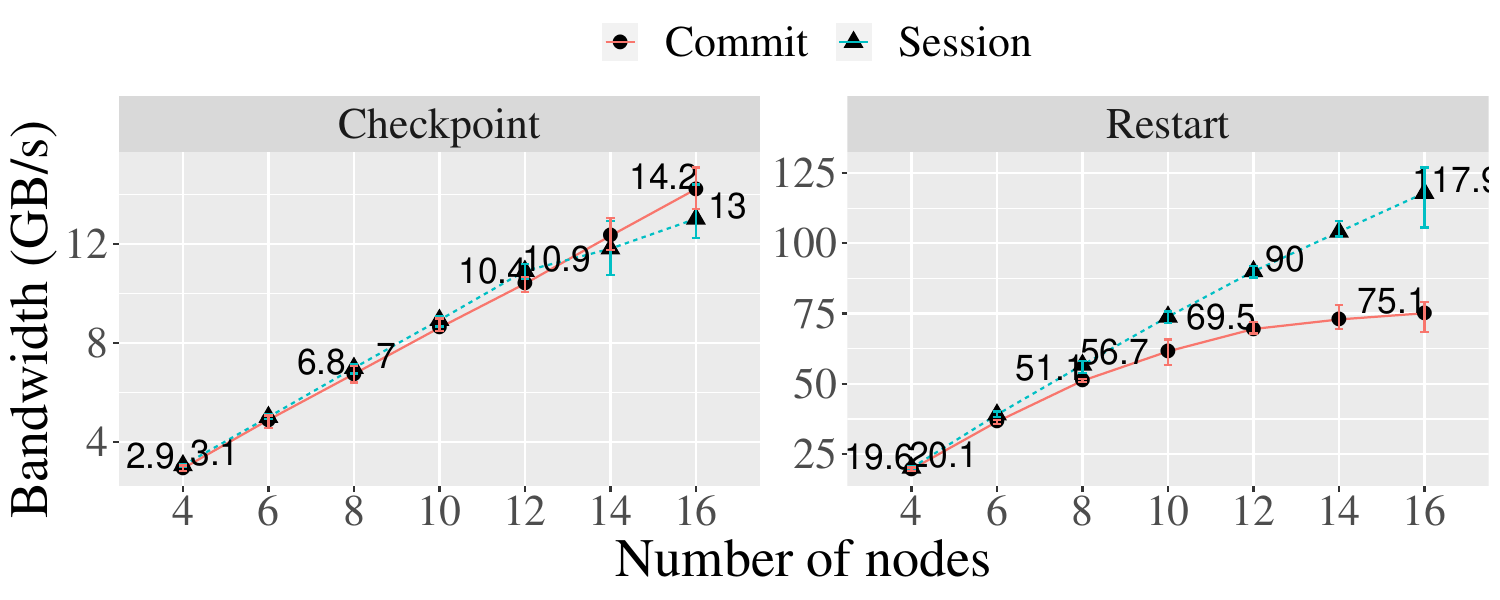}
    \caption{HACC-IO with SCR.}
    \label{fig:iobench_haccio_scr}
\end{figure}

We show the read and write bandwidths of checkpoint and restart phases in Figure~\ref{fig:iobench_haccio_scr}. To better understand the read bandwidth, the result did not include the communication time for the spare node to get the checkpoint. Similar to the large-write experiments discussed earlier, SCR scaled well for checkpointing and achieved the peak bandwidth at all scales under both consistency models. In other words, the consistency model does not have a big impact on SCR's checkpointing bandwidth. However, for restarting, session consistency scaled better than commit consistency, mainly due to the low query frequency. At the restart phase, the reads were satisfied through memory buffers, and the overall read bandwidth scaled linearly with the number of nodes, which made the read time per node constant. However, under commit consistency, when more nodes were used, more query requests (one per read) were sent simultaneously to the global server, which then became the bottleneck and reduced scalability.

\subsection{Case Study: I/O of Distributed Deep Learning}

Deep learning has thrived in recent years. However, as data sizes and system scales increase, traditional methods of feeding neural networks during training struggle to keep up with the required computation. To accelerate data ingestion rates, various methods~\cite{dl_hybrid_parallelism, dl_large_minibatch, lbann_parallel, LTFB} have been proposed, such as data sharding, prestaging, and in-memory caching.

Here, we simulate the I/O of the ``Preloaded'' strategy that was proposed in~\cite{lbann_parallel} and implemented in LBANN~\cite{app_lbann}. Our simulation assigns to each process a non-overlapping subset of the training data. Before the training begins, each process loads its portion of data into its node-local SSD (hence the term Preloaded).
Next, at the beginning of each epoch, each process is assigned a random subset of samples. The samples are evenly distributed to all processes so that each process performs an equal amount of work. During each epoch, each process reads the assigned samples, either locally or from other processes using MPI. It is worth noting that our benchmark is a simplified version of the Preloaded strategy that differs in two major ways: (1) we store data in node-local SSDs instead of memory, which is anyhow necessary for large datasets that do not fit in memory; and (2) we do not perform aggregations when sending samples to the same process, which places additional stress on the file system.

\begin{figure}[htbp]
    \centering
    \includegraphics[width=\linewidth]{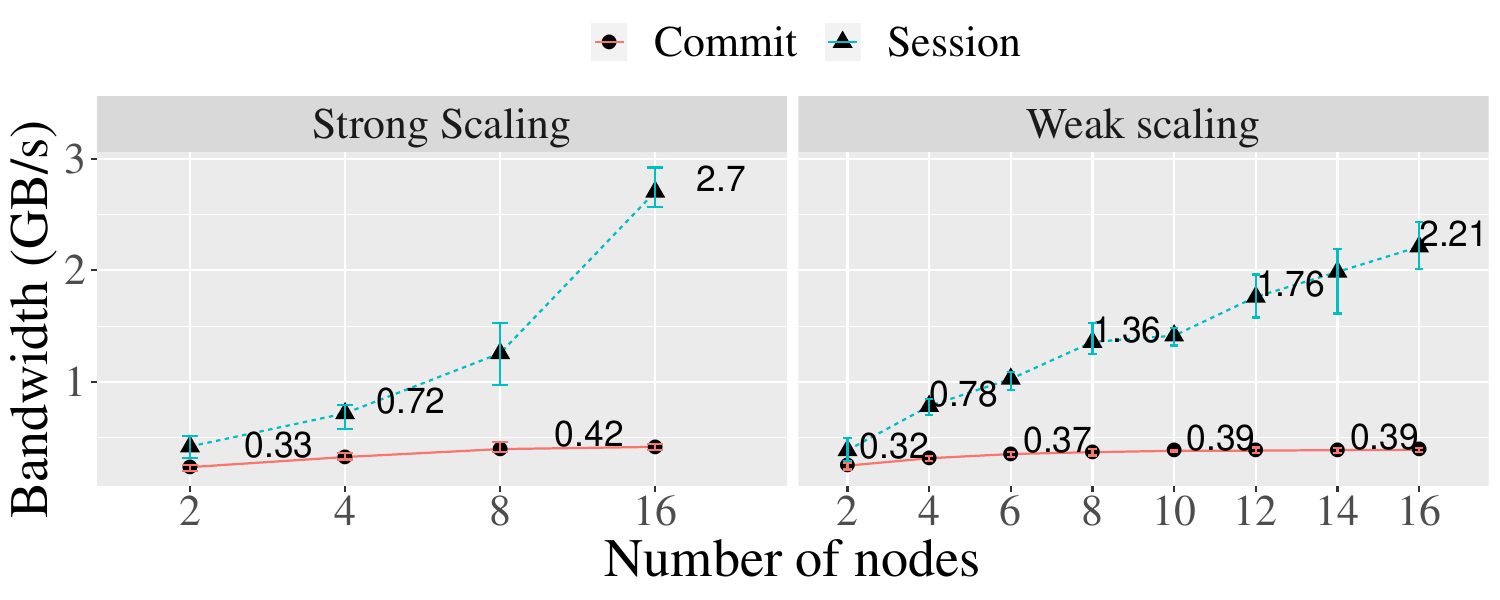}
    \caption{Random read bandwidth of DL application.}
    \label{fig:iobench_dl}
\end{figure}

The average per-epoch read bandwidth is presented in Figure~\ref{fig:iobench_dl}. We conducted both strong scaling and weak scaling experiments, with a mini-batch size of 1024 for strong scaling, and each process working on 32 samples per iteration for weak scaling. The sample size was set to 116KB, which is the same as the average image size of ImageNet-1K~\cite{ImageNet}. The number of processes per node was set to 4 (in real DL training, this number is usually set to match the number of GPUs per node).
The results are very similar to those of small-reads experiments shown in Figure~\ref{fig:iobench_ccr_csr_10x8kb}, only the bandwidth is higher here thanks to the slightly larger reads (116KB vs. 8KB). In both strong scaling and weak scaling, session consistency outperformed commit consistency in terms of scalability and bandwidth, due to the less time spent on queries. Additionally, the increasing gap in bandwidth between the two consistency models with the number of nodes further emphasizes the significance of choosing an appropriate consistency model to achieve optimal performance and scalability.

\subsection{Key Takeaways}

Here, we present the key findings derived from our experiments.

\begin{itemize}
    \item When performing large writes and reads (e.g., over one megabyte per I/O operation), consistency models do not have a big impact on the I/O bandwidth. This is because the overhead of maintaining the consistency model (weaker or stronger) is insignificant compared to the time needed to to access the I/O device.
    \item When performing small writes and reads (e.g., ranging from a few bytes to a few kilobytes), the adoption of a stronger consistency model can noticeably hinder performance. This is attributed to the faster completion of each I/O operation, making the overhead of maintaining strong consistency a bottleneck. Moreover, the traffic required to maintain the consistency model can lead to contention, particularly when there is a high volume of small I/O operations.
    \item When I/O operations are directly fulfilled by memory or fast devices like persistent memory, the choice of consistency models can significantly impact performance. This is due to a similar reason as mentioned earlier, where the faster completion of I/O operations magnifies the overhead associated with maintaining strong consistency models.
    \item For small random accesses (e.g., random reads of deep learning applications), weaker consistency models demonstrate higher I/O bandwidth and improved scalability compared to stronger models. Notably, this improvement is significant even at smaller scales, indicating a promising direction for optimizing the I/O performance of deep learning applications.
\end{itemize}

\section{Conclusion and Future Work}
\label{sec:conclusion}

This work explored consistency models from the perspective of parallel file systems. We provided a high-level discussion on important aspects of storage consistency models, including their design choices and their comparison with memory models. Based on the commonalities of existing storage models, we proposed a unified and formal framework for specifying properly-synchronized SCNF models, which guarantee sequential consistency (or POSIX consistency) for programs that are properly synchronized. Additionally, we proposed a flexible design for implementing properly-synchronized SCNF models that isolates the consistency model from other file system components, making it easy to understand the impact of different consistency models on I/O performance.

We also presented a detailed performance comparison between commit consistency and session consistency. Our results indicate that session consistency is better suited for most HPC I/O workloads in terms of performance and scalability. Although this comes at the cost of slightly reduced programmability, the performance gain is potentially huge, especially for small reads such as those in deep learning applications. Overall, this work contributes to a better understanding of consistency models in parallel file systems and their impact on I/O performance.

In our future work, we will implement different relaxed storage models in existing PFSs to evaluate their performance impacts in a real-world setting. Additionally, we plan to study the consistency requirements of metadata operations for HPC applications and evaluate their performance implications.

\section*{Acknowledgments}
\noindent This work was supported by NSF SHF Collaborative grant 1763540 and was performed under the auspices of the U.S. Department of Energy by Lawrence Livermore National Laboratory under Contract DE-AC52-07NA27344. LLNL-JRNL-849174-DRAFT. This material is based upon work supported by the U.S. Department of Energy, Office of Science, Office of Advanced Scientific Computing Research under the DOE Early Career Research Program.

\bibliography{main}

\begin{thebibliography}{10}
\providecommand{\url}[1]{#1}
\csname url@samestyle\endcsname
\providecommand{\newblock}{\relax}
\providecommand{\bibinfo}[2]{#2}
\providecommand{\BIBentrySTDinterwordspacing}{\spaceskip=0pt\relax}
\providecommand{\BIBentryALTinterwordstretchfactor}{4}
\providecommand{\BIBentryALTinterwordspacing}{\spaceskip=\fontdimen2\font plus
\BIBentryALTinterwordstretchfactor\fontdimen3\font minus
  \fontdimen4\font\relax}
\providecommand{\BIBforeignlanguage}[2]{{%
\expandafter\ifx\csname l@#1\endcsname\relax
\typeout{** WARNING: IEEEtran.bst: No hyphenation pattern has been}%
\typeout{** loaded for the language `#1'. Using the pattern for}%
\typeout{** the default language instead.}%
\else
\language=\csname l@#1\endcsname
\fi
#2}}
\providecommand{\BIBdecl}{\relax}
\BIBdecl

\bibitem{patel2019revisiting}
T.~Patel, S.~Byna, G.~K. Lockwood, and D.~Tiwari, ``{Revisiting I/O Behavior in
  Large-Scale Storage Systems: the Expected and the Unexpected},'' in
  \emph{Proceedings of the International Conference for High Performance
  Computing, Networking, Storage and Analysis}, 2019, pp. 1--13.

\bibitem{paul2020understanding}
A.~K. Paul, O.~Faaland, A.~Moody, E.~Gonsiorowski, K.~Mohror, and A.~R. Butt,
  ``{Understanding HPC Application I/O Behavior Using System Level
  Statistics},'' in \emph{2020 IEEE 27th International Conference on High
  Performance Computing, Data, and Analytics (HiPC)}.\hskip 1em plus 0.5em
  minus 0.4em\relax IEEE, 2020, pp. 202--211.

\bibitem{clairvoyant}
N.~Dryden, R.~B{\"o}hringer, T.~Ben-Nun, and T.~Hoefler, ``{Clairvoyant
  Prefetching for Distributed Machine Learning I/O},'' \emph{arXiv preprint
  arXiv:2101.08734}, 2021.

\bibitem{mummi}
F.~Di~Natale, H.~Bhatia, T.~S. Carpenter, C.~Neale, S.~Kokkila-Schumacher,
  T.~Oppelstrup, L.~Stanton, X.~Zhang, S.~Sundram, T.~R. Scogland
  \emph{et~al.}, ``{A Massively Parallel Infrastructure for Adaptive Multiscale
  Simulations: Modeling RAS Initiation Pathway for Cancer},'' in
  \emph{Proceedings of the International Conference for High Performance
  Computing, Networking, Storage and Analysis}, 2019, pp. 1--16.

\bibitem{POSIX}
``{IEEE Standard for Information Technology--Portable Operating System
  Interface (POSIX(TM)) Base Specifications, Issue 7},'' \emph{{IEEE Std
  1003.1-2017 (Revision of IEEE Std 1003.1-2008)}}, pp. 1--3951, 2018.

\bibitem{Lustre}
P.~Braam, ``{The Lustre Storage Architecture},'' \emph{arXiv preprint
  arXiv:1903.01955}, 2019.

\bibitem{GPFS}
F.~B. Schmuck and R.~L. Haskin, ``{GPFS: A Shared-Disk File System for Large
  Computing Clusters.}'' in \emph{FAST}, vol.~2, no.~19, 2002.

\bibitem{BeeGFS}
\BIBentryALTinterwordspacing
F.~Herold, S.~Breuner, and J.~Heichler, ``{An Introduction to BeeGFS},'' 2014.
  [Online]. Available:
  \url{https://www.beegfs.io/docs/whitepapers/Introduction_to_BeeGFS_by_ThinkParQ.pdf}
\BIBentrySTDinterwordspacing

\bibitem{BurstFS}
T.~Wang, K.~Mohror, A.~Moody, W.~Yu, and K.~Sato, ``{BurstFS: A Distributed
  Burst Buffer File System for Scientific Applications},'' in \emph{The
  International Conference for High Performance Computing, Networking, Storage
  and Analysis (SC)}, 2015.

\bibitem{UnifyFS}
L.~L.~N. Laboratory, ``{UnifyFS: A File System for Burst Buffers },''
  \url{https://github.com/LLNL/UnifyFS}, Mar. 2021.

\bibitem{echofs}
A.~Miranda, R.~Nou, and T.~Cortes, ``{echofs: A Scheduler-Guided Temporary
  Filesystem to Leverage Node-local NVMs},'' in \emph{2018 30th International
  Symposium on Computer Architecture and High Performance Computing
  (SBAC-PAD)}.\hskip 1em plus 0.5em minus 0.4em\relax IEEE, 2018, pp. 225--228.

\bibitem{GfarmBB}
O.~Tatebe, S.~Moriwake, and Y.~Oyama, ``{Gfarm/BB—Gfarm File System for
  Node-Local Burst Buffer},'' \emph{Journal of Computer Science and
  Technology}, vol.~35, no.~1, pp. 61--71, 2020.

\bibitem{SymphonyFS}
S.~Oral, S.~S. Vazhkudai, F.~Wang, C.~Zimmer, C.~Brumgard, J.~Hanley,
  G.~Markomanolis, R.~Miller, D.~Leverman, S.~Atchley \emph{et~al.},
  ``{End-to-end I/O Portfolio for the Summit Supercomputing Ecosystem},'' in
  \emph{Proceedings of the International Conference for High Performance
  Computing, Networking, Storage and Analysis}, 2019, pp. 1--14.

\bibitem{lamport_sc}
L.~Lamport, ``{How to Make a Multiprocessor Computer That Correctly Executes
  Multiprocess Progranm},'' \emph{IEEE transactions on computers}, vol.~28,
  no.~09, pp. 690--691, 1979.

\bibitem{TSO}
P.~Sewell, S.~Sarkar, S.~Owens, F.~Z. Nardelli, and M.~O. Myreen, ``{x86-TSO: A
  Rigorous and Usable Programmer's Model for x86 Multiprocessors},''
  \emph{Communications of the ACM}, vol.~53, no.~7, pp. 89--97, 2010.

\bibitem{weak_ordering}
M.~Dubois, C.~Scheurich, and F.~Briggs, ``{Memory Access Buffering in
  Multiprocessors},'' \emph{ACM SIGARCH computer architecture news}, vol.~14,
  no.~2, pp. 434--442, 1986.

\bibitem{release_consistency}
K.~Gharachorloo, D.~Lenoski, J.~Laudon, P.~Gibbons, A.~Gupta, and J.~Hennessy,
  ``{Memory Consistency and Event Ordering in Scalable Shared-Memory
  Multiprocessors},'' \emph{ACM SIGARCH Computer Architecture News}, vol.~18,
  no. 2SI, pp. 15--26, 1990.

\bibitem{wang2020hpdc}
C.~Wang, K.~Mohror, and M.~Snir, ``{File System Semantics Requirements of HPC
  Applications},'' in \emph{Proceedings of the 30th International Symposium on
  High-Performance Parallel and Distributed Computing (HPDC)}, 2020, pp.
  19--30.

\bibitem{BSCFS}
\BIBentryALTinterwordspacing
IBM, ``{Burst Buffer Shared Checkpoint File System},'' Apr. 2020. [Online].
  Available: \url{https://github.com/IBM/CAST/tree/master/bscfs}
\BIBentrySTDinterwordspacing

\bibitem{NFSv4}
S.~Shepler, B.~Callaghan, D.~Robinson, R.~Thurlow, C.~Beame, M.~Eisler, and
  D.~Noveck, ``{RFC3530: Network File System (NFS) Version 4 Protocol},'' 2003.

\bibitem{MPI-IO}
P.~Corbett, D.~Feitelson, S.~Fineberg, Y.~Hsu, B.~Nitzberg, J.-P. Prost,
  M.~Snir, B.~Traversat, and P.~Wong, ``{Overview of the MPI-IO Parallel I/O
  Interface},'' in \emph{IPPS’95 Workshop on Input/Output in Parallel and
  Distributed Systems}, 1995, pp. 1--15.

\bibitem{MPI4}
``{MPI: A Message-Passing Interface Standard Version 4.0},''
  \url{https://www.mpi-forum.org/docs/mpi-4.0/mpi40-report.pdf}, 2021.

\bibitem{adve1993thesis}
S.~V. Adve, ``{Designing Memory Consistency Models for Shared-Memory
  Multiprocessors},'' Ph.D. dissertation, University of Wisconsin, Madison,
  1993.

\bibitem{adve1990DRF0}
S.~V. Adve and M.~D. Hill, ``{Weak Ordering - a New Definition},'' \emph{ACM
  SIGARCH Computer Architecture News}, vol.~18, no. 2SI, pp. 2--14, 1990.

\bibitem{java_memory_model}
J.~Manson, W.~Pugh, and S.~V. Adve, ``{The Java Memory Model},'' \emph{ACM
  SIGPLAN Notices}, vol.~40, no.~1, pp. 378--391, 2005.

\bibitem{SCR}
A.~Moody, G.~Bronevetsky, K.~Mohror, and B.~R. De~Supinski, ``{Design,
  Modeling, and Evaluation of a Scalable Multi-level Checkpointing System},''
  in \emph{SC'10: Proceedings of the 2010 ACM/IEEE International Conference for
  High Performance Computing, Networking, Storage and Analysis}.\hskip 1em plus
  0.5em minus 0.4em\relax IEEE, 2010, pp. 1--11.

\bibitem{hacc-io}
``{HACC IO Kernel from the CORAL Benchmark Codes},''
  \url{https://asc.llnl.gov/coral-benchmarks##hacc}, Jan 2018.

\bibitem{dl_hybrid_parallelism}
Y.~Oyama, N.~Maruyama, N.~Dryden, E.~McCarthy, P.~Harrington, J.~Balewski,
  S.~Matsuoka, P.~Nugent, and B.~Van~Essen, ``{The Case for Strong Scaling in
  Deep Learning: Training Large 3D CNNs With Hybrid Parallelism},'' \emph{IEEE
  Transactions on Parallel and Distributed Systems}, vol.~32, no.~7, pp.
  1641--1652, 2020.

\bibitem{dl_large_minibatch}
P.~Goyal, P.~Doll{\'a}r, R.~Girshick, P.~Noordhuis, L.~Wesolowski, A.~Kyrola,
  A.~Tulloch, Y.~Jia, and K.~He, ``{Accurate, Large Minibatch SGD: Training
  ImageNet in 1 Hour},'' \emph{arXiv preprint arXiv:1706.02677}, 2017.

\bibitem{lbann_parallel}
S.~A. Jacobs, B.~Van~Essen, D.~Hysom, J.-S. Yeom, T.~Moon, R.~Anirudh, J.~J.
  Thiagaranjan, S.~Liu, P.-T. Bremer, J.~Gaffney \emph{et~al.},
  ``{Parallelizing Training of Deep Generative Models on Massive Scientific
  Datasets},'' in \emph{2019 IEEE International Conference on Cluster Computing
  (CLUSTER)}.\hskip 1em plus 0.5em minus 0.4em\relax IEEE, 2019, pp. 1--10.

\bibitem{LTFB}
S.~A. Jacobs, N.~Dryden, R.~Pearce, and B.~Van~Essen, ``{Towards Scalable
  Parallel Training of Deep Neural Networks},'' in \emph{Proceedings of the
  Machine Learning on HPC Environments}, 2017, pp. 1--9.

\bibitem{app_lbann}
\BIBentryALTinterwordspacing
B.~Van~Essen, H.~Kim, R.~Pearce, K.~Boakye, and B.~Chen, ``{LBANN: Livermore
  Big Artificial Neural Network HPC Toolkit},'' in \emph{Proceedings of the
  Workshop on Machine Learning in High-Performance Computing Environments},
  ser. MLHPC '15.\hskip 1em plus 0.5em minus 0.4em\relax New York, NY, USA:
  ACM, 2015, pp. 5:1--5:6. [Online]. Available:
  \url{http://doi.acm.org/10.1145/2834892.2834897}
\BIBentrySTDinterwordspacing

\bibitem{ImageNet}
J.~Deng, W.~Dong, R.~Socher, L.-J. Li, K.~Li, and L.~Fei-Fei, ``Imagenet: A
  large-scale hierarchical image database,'' in \emph{2009 IEEE conference on
  computer vision and pattern recognition}.\hskip 1em plus 0.5em minus
  0.4em\relax Ieee, 2009, pp. 248--255.

\end{thebibliography}
\bibliographystyle{IEEEtran}

\begin{IEEEbiography}[{\includegraphics[width=1in,clip,keepaspectratio]{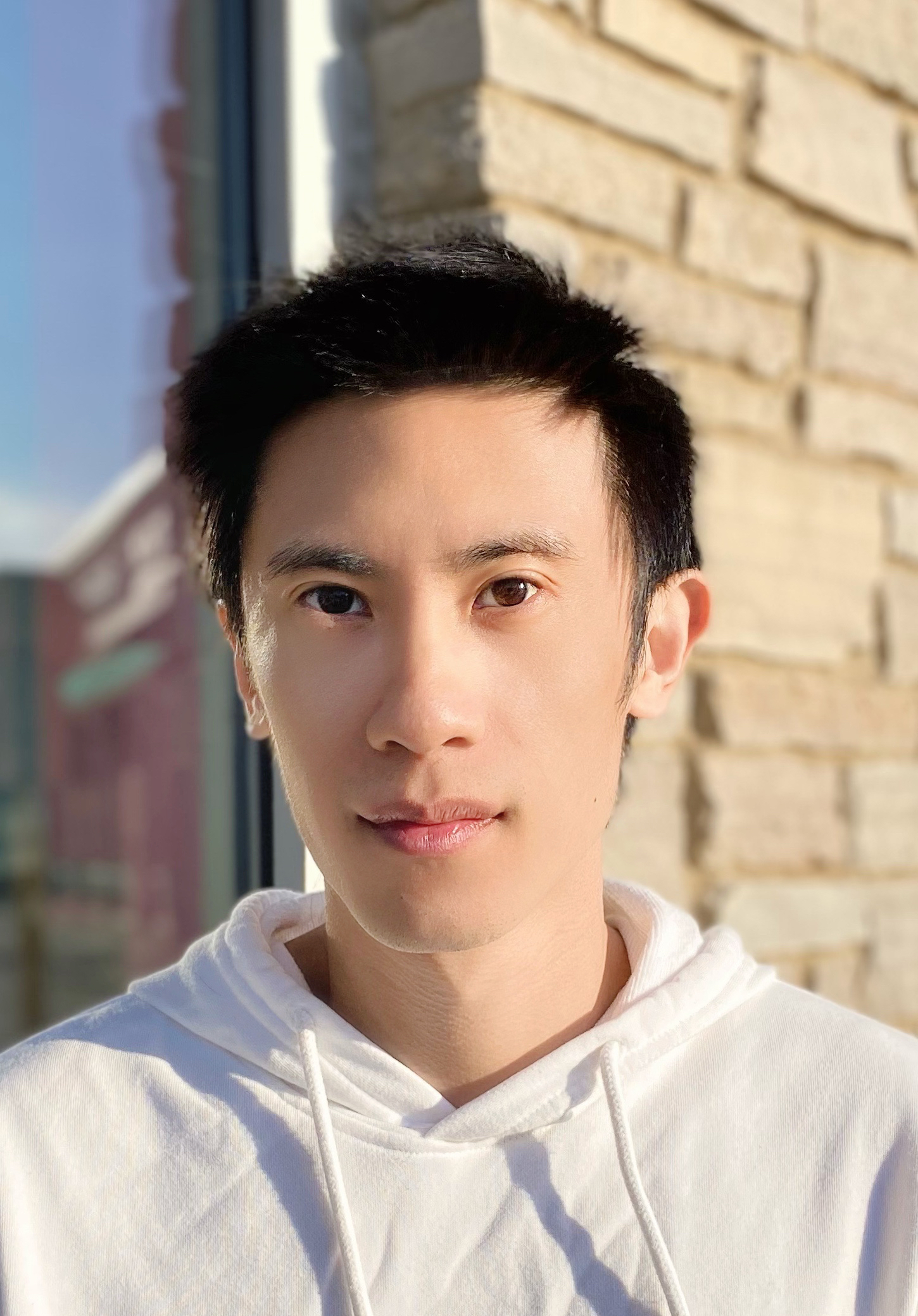}}]{Chen Wang}
is Fernbach Postdoctoral Fellow at Lawrence Livermore National Laboratory. He is currently working at the Center for Applied Scientific Computing at LLNL.
He received his Ph.D in computer science from University of Illinois Urbana-Champaign. His research interests include parallel computing, I/O and communication tracing, and parallel storage systems.
\end{IEEEbiography}
\begin{IEEEbiography}[{\includegraphics[width=1in,clip,keepaspectratio]{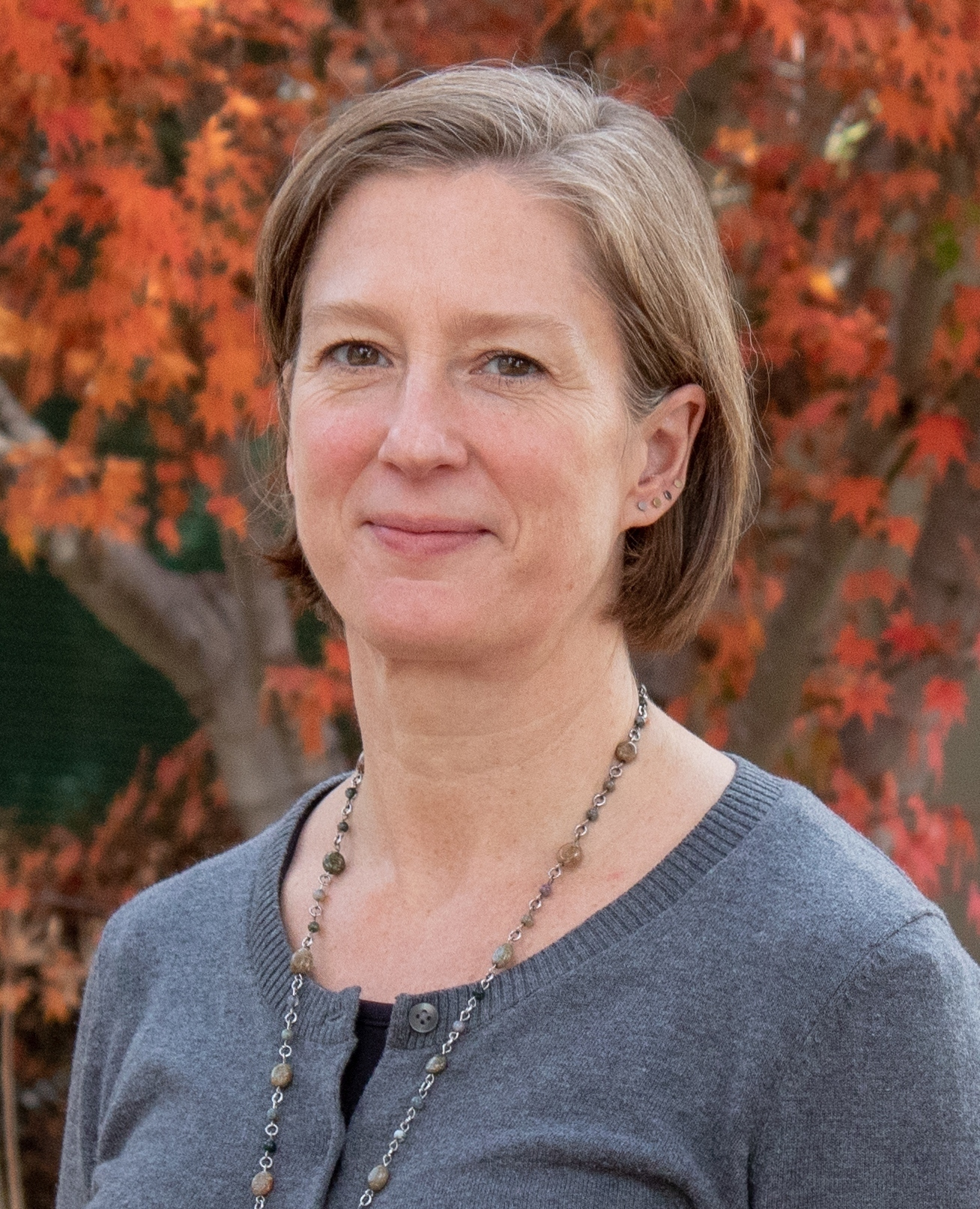}}]{Kathryn Mohror}
is a computer scientist in the Parallel Systems Group in the Center for Applied Scientific Computing (CASC) at Lawrence Livermore National Laboratory (LLNL). Kathryn serves as the Deputy Director for the Laboratory Directed Research \& Development (LDRD) program at LLNL, Lead for the NNSA Software Technologies Portfolio for the U.S. Exascale Computing Project (ECP), and as the ASCR Point of Contact for Computer Science at LLNL. Kathryn’s research on high-end computing systems is currently focused on I/O for extreme scale systems. Her other research interests include scalable performance analysis and tuning, fault tolerance, and parallel programming paradigms. 
\end{IEEEbiography}
\begin{IEEEbiography}[{\includegraphics[width=1in,clip,keepaspectratio]{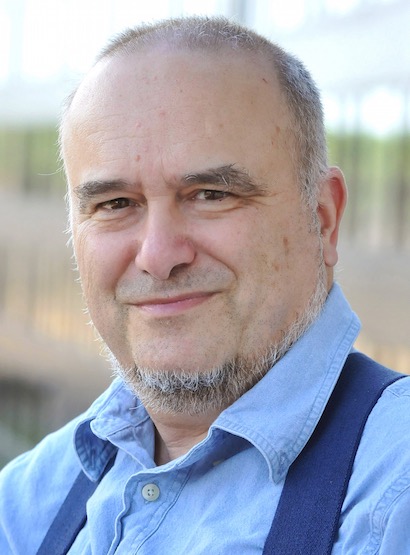}}]{Marc Snir}
is Michael Faiman Emeritus Professor in the Department of Computer Science at the University of Illinois Urbana-Champaign. He was Director of the Mathematics and Computer Science Division at the Argonne National Laboratory from 2011 to 2016 and head of the Computer Science Department at Illinois from 2001 to 2007. Until  2001 he was a senior manager at the IBM T. J. Watson Research Center where he led the Scalable Parallel Systems research group that was responsible for major contributions to the IBM scalable parallel system and to the IBM Blue Gene system.
\end{IEEEbiography}

\end{document}